\documentclass
[superscriptaddress,secnumarabic,amssymb,amsmath,nobibnotes,aps,prd,showkeys,showpacs,nofootinbib,onecolumn]{revtex4}%
\usepackage{amsmath}
\usepackage{graphicx}
\usepackage{epstopdf}
\usepackage{epsf}
\usepackage{eurosym}
\usepackage{amsfonts}
\usepackage{dcolumn}
\usepackage{setspace}
\usepackage{color}
\usepackage{verbatim}
\usepackage{graphicx}
\usepackage{caption}
\usepackage{graphics}
\usepackage{slashed}
\usepackage{amssymb}%
\providecommand{\U}[1]{\protect\rule{.1in}{.1in}}
\definecolor{darkgreen}{rgb}{0,0.35,0}

\newcommand{\be}{\begin{equation}}
\newcommand{\ee}{\end{equation}}

\newcommand{\mincir}{\raise
-3.truept\hbox{\rlap{\hbox{$\sim$}}\raise4.truept\hbox{$<$}\ }}
\newcommand{\magcir}{\raise
-3.truept\hbox{\rlap{\hbox{$\sim$}}\raise4.truept\hbox{$>$}\ }}

\ifx\pdfoutput\relax\let\pdfoutput=\undefined\fi
\newcount\msipdfoutput
\ifx\pdfoutput\undefined\else
\ifcase\pdfoutput\else
\ifx\paperwidth\undefined\else
\ifdim\paperheight=0pt\relax\else\pdfpageheight\paperheight\fi
\ifdim\paperwidth=0pt\relax\else\pdfpagewidth\paperwidth\fi
\fi\fi\fi
\begin{document}
\title{Cosmological Einstein-Skyrme solutions with non-vanishing topological charge}
\author{Fabrizio Canfora}
\email{canfora-at-cecs.cl}
\affiliation{Centro de Estudios Cient\'{\i}ficos (CECS), Arturo Prat 514, Valdivia, Chile}
\author{Andronikos Paliathanasis}
\email{anpaliat@phys.uoa.gr}
\affiliation{Instituto de Ciencias F\'{\i}sicas y Matem\'{a}ticas, Universidad Austral de
Chile, Valdivia, Chile}
\affiliation{Institute of Systems Science, Durban University of Technology, Durban 4000,
South Africa}
\author{Tim Taves}
\email{timtaves@gmail.com}
\affiliation{Centro de Estudios Cient\'{\i}ficos (CECS), Arturo Prat 514, Valdivia, Chile}
\author{Jorge Zanelli}
\email{z-at-cecs.cl}
\affiliation{Centro de Estudios Cient\'{\i}ficos (CECS), Arturo Prat 514, Valdivia, Chile}

\begin{abstract}
Time-dependent analytic solutions of the Einstein-Skyrme system --gravitating
Skyrmions--, with topological charge one are analyzed in detail. In
particular, the question of whether these Skyrmions reach a spherically
symmetric configuration for $t\rightarrow+\infty$ is discussed. It is shown
that there is a static, spherically symmetric solution described by the
Ermakov-Pinney system, which is fully integrable by algebraic methods. For
$\Lambda>0$ this spherically symmetric solution is found to be in a ``neutral
equilibrium" under small deformations, in the sense that under a small
squashing it would neither blow up nor dissapear after a long time, but it
would remain finite forever (plastic deformation). Thus, in a sense, the
coupling with Einstein gravity spontaneously breaks the spherical symmetry of
the solution. However, in spite of the lack of isotropy, for $t \to\infty$
(and $\Lambda>0$) the space time is locally flat and the anisotropy of the
Skyrmion only reflects the squashing of spacetime.

\end{abstract}
\maketitle


\section{Introduction}


The Skyrme system \cite{skyrme} is one of the most useful models in nuclear
and particle physics due to its close relationship to low energy QCD
\cite{witten0}. A remarkable feature of the Skyrme action is that it allows
for the existence of solitons (Skyrmions) that behave as Fermionic degrees of
freedom, in spite of the fact that the basic fields are scalar. Furthermore,
Skyrmions describe nucleons both theoretically and phenomenologically (see, e.
g.,\cite{witten0,finkrub,multis2,manton,susy,giulini,bala2,bala0,bala1,ANW,guada,rec1,rec2}%
), where the identification of the winding number of the Skyrmion with the
Baryon number in particle physics \cite{witten0} plays a crucial role.
Following \cite{finkrub,giulini}, the possibility of treating the Skyrme
solitons as Fermions was extended to curved spaces as well
\cite{curved1f,curved2f}, opening the possibility for applying this theory to
general relativity and astrophysics.

The above reasons imply that the Einstein-Skyrme system might be relevant for
astrophysics from a phenomenological point of view. From a more theoretical
angle, numerical computations following earlier results in \cite{lucock,droz}
indicate the existence of spherically symmetric black-hole solutions with a
nontrivial Skyrme field (Skyrme hair) \cite{bh01,bh02}. These were the first
counterexamples to the black hole no-hair conjecture, and, moreover, the
stability against spherical linear perturbations was shown in \cite{droz2}.
Regular particle-like configurations \cite{numerical1} and dynamical
properties of the system have also been investigated numerically
\cite{numerical2}. Even in the sector with vanishing topological charge the
cosmological consequences of the Skyrme model are quite interesting
\cite{cosmo,cosmo2,cosmo3}.

Thus, having analytic solutions of the Einstein-Skyrme system with nontrivial
topological charges would be extremely useful. In particular, the
gravitational implications of the discreteness of the topological charge
together with the fact that such topological objects have a characteristic
size, deserve an in-depth investigation. An especially compelling case is the
time-dependent situation in which the coupling of the Skyrme system with
gravity could reveal unexpected departures from the ``natural" spherical
symmetry of configurations with winding number $W=1$.

At first glance, the possibility of finding nontrivial analytic solutions of
the Einstein-Skyrme system may seem hopeless. Until a few years ago, no
analytic solutions of the Skyrme model in flat space had been found. Quite
recently, however, the generalized hedgehog ansatz (introduced in
\cite{canfora} and its generalizations in
\cite{canfora2,canfora3,canfora4,yang1,canfora5,yang2,canfora6,canfora7})
allowed for the construction of exact multi-Skyrmion configurations as well as
the first analytic gravitating Skyrmions \cite{canfora6}. Moreover, these
approaches also work in the Yang-Mills case \cite{canfora7}\textbf{.}

In \cite{canfora6}, the full Einstein-Skyrme field equations, in the Bianchi
IX case and the $W=1$ sector, reduce to a system of two autonomous second
order ODEs for two scale factors, where the Skyrme field equations, which are
usually the difficult part of the problem, are automatically satisfied in this
ansatz. Such a system allows addressing the question of whether or not the
$W=1$ Skyrmion --which is known to be spherically symmetric in flat space--
retains this symmetry when coupled to gravity. A preliminary analysis in the
reference \cite{sergei} suggests that the answer should depend on the value of
the cosmological constant and not just on its sign. Here we generalize the
analysis of \cite{sergei}, confirming that the cosmological constant is one of
the relevant parameters of the dynamical evolution. Moreover, we also clarify
in which sense the $t\rightarrow\infty$ evolution of the system is ``
asymptotically" spherically symmetric, that is, ``asymptotically isotropic".
This paper is organized as follows.

The action integral for the Einstein-Skyrme model with cosmological constant
is presented in Section \ref{section2} where we introduce the self-gravitating
Skyrmion model in the background geometry of a locally rotational Bianchi IX
universe. The remaining equations are those of General Relativity in which the
energy momentum tensor is produced by the Skyrmion. In Section \ref{section3}
the field equations are shown to describe a mechanical system of two degrees
of freedom. In the limit in which the Bianchi IX space-time is isotropic
corresponding to an Einstein-Skyrme system with $W=1$, the integrability of
the Ermakov-Pinney system provides a solution, including a special solutions
for the static Einstein universe. In Section \ref{isotro} we analyze the
stability of the isotropic $W=1$ solution by studying the first-order
perturbations around it and show that it is not stable. However for a positive
cosmological constant we show that the final universe is approximately
isotropic. The discussion of our results and our conclusions are given in
Section \ref{conc}.


\section{The Action Integral}

\label{section2}

We are interested in self-gravitating Skyrmions for the $SU(2)$ group
described by the action
\begin{equation}
I[g,U]=\int d^{4}x\sqrt{-g}\left(  \frac{R-2\Lambda}{2\kappa}+\frac{K}%
{4}\mathrm{Tr}[A^{\mu}A_{\mu}+\frac{\lambda}{8}F_{\mu\nu}F^{\mu\nu}]\right)  .
\label{skyrmaction}%
\end{equation}
Here $A_{\mu}$ is a shorthand for the Maurer-Cartan form $U^{-1}\nabla_{\mu}%
U$, with $U\in SU(2)$ and $F_{\mu\nu}=\left[  A_{\mu},A_{\nu}\right]  $;
$A_{\mu}=A_{\mu}^{j}t_{j}$ where $t_{j}=-i\sigma_{j}$ are the $SU(2)$
generators, and $\sigma_{j}$ are the Pauli matrices. In our conventions
$c=\hbar=1$, the spacetime signature is $(-,+,+,+)$ and Greek indices run over
spacetime. Moreover, $R$ is the Ricci scalar, $\Lambda$ is the cosmological
constant and $\kappa$ is the gravitational constant. Here $K$ and $\lambda$
are (positive) coupling constants, related to the experimentally determined
phenomenological parameters $F_{\pi}$ and $e$ through \cite{ANW}
\begin{align*}
K  &  =\frac{1}{4}F_{\pi}^{2},\quad K\lambda=\frac{1}{e^{2}}\ ,\\
F_{\pi}  &  =186\ \ MeV\ ,\ e=5.45\ .
\end{align*}

The Skyrme equation, obtained by varying (\ref{skyrmaction}) with respect to
$U$, together with Einstein's equations are
\begin{subequations}
\label{system}%
\begin{align}
\nabla^{\mu}A_{\mu}+\frac{\lambda}{4}\nabla^{\mu}[A^{\nu},F_{\mu\nu}]  &
=0\ ,\label{nonlinearsigma1}\\
G_{\mu\nu}+\Lambda g_{\mu\nu}  &  =\kappa T_{\mu\nu}, \label{einstein}%
\end{align}
where $G_{\mu\nu}$ is the Einstein tensor and the energy-momentum tensor for
the Skyrme field is%

\end{subequations}
\begin{equation}
T_{\mu\nu}=-\frac{K}{2}\mathrm{Tr}\left[  A_{\mu}A_{\nu}-\frac{1}{2}g_{\mu\nu
}A^{\alpha}A_{\alpha}+\frac{\lambda}{4}\left(  g^{\alpha\beta}F_{\mu\alpha
}F_{\nu\beta}-\frac{1}{4}g_{\mu\nu}F_{\alpha\beta}F^{\alpha\beta}\right)
\right]  . \label{tmunu}%
\end{equation}


\subsection{Static self-gravitating Skyrmion}


The spacetime geometry for the static solutions of the coupled system
(\ref{system}) is the product $R\times S^{3}$,
\begin{equation}
ds^{2}=-dt^{2}+\frac{\rho_{0}^{2}}{4}\left[  (d\gamma+\cos\theta d\varphi
)^{2}+d\theta^{2}+\sin^{2}\theta d\varphi^{2}\right]  , \label{metric1}%
\end{equation}
where $0\leq\gamma\leq2\pi$, $0\leq\theta\leq\pi$, $0\leq\varphi\leq2\pi$ are
the coordinates on the 3-sphere of constant radius $\rho_{0}$.

Following \cite{canfora,canfora2,canfora3,canfora4}, \cite{canfora5},
\cite{canfora6}, we adopt the standard parametrization of the $SU(2)$-valued
scalar $U(x^{\mu})$ as
\begin{equation}
U^{\pm1}(x^{\mu})=Y^{0}(x^{\mu})\pmb{\mathbb{I}}\pm Y^{i}(x^{\mu}%
)t_{i},\ \ \left(  Y^{0}\right)  ^{2}+Y^{i}Y_{i}=1, \label{standard1}%
\end{equation}
where $\pmb{\mathbb{I}}$ is the $2\times2$ identity matrix. The unit vector
$Y^{A}=(Y^{0},Y^{i})$ defines the embedded three sphere, which is naturally
given by
\begin{subequations}
\label{3-sphere}%
\begin{align}
Y^{0}  &  =\cos\alpha, & Y^{i}  &  =n^{i}\sin\alpha, &  & \label{pions1}\\
n^{1}  &  =\sin\Theta\cos\Phi, & n^{2}  &  =\sin\Theta\sin\Phi, & n^{3}  &
=\cos\Theta. \label{pions2}%
\end{align}
With this information one can solve (\ref{nonlinearsigma1}) for $\alpha$,
$\Theta$ and $\Phi$ as functions of $\gamma$, $\theta$ and $\varphi$. It can
be directly checked that the configuration
\end{subequations}
\begin{equation}
\Phi=\frac{\gamma+\varphi}{2},\ \tan\Theta=\frac{\cot\left(  \frac{\theta}%
{2}\right)  }{\cos\left(  \frac{\gamma-\varphi}{2}\right)  },\ \tan
\alpha=\frac{\sqrt{1+\tan^{2}\Theta}}{\tan\left(  \frac{\gamma-\varphi}%
{2}\right)  }, \label{pions2.25}%
\end{equation}
identically satisfies the Skyrme equations (\ref{nonlinearsigma1}) in the
background metric (\ref{metric1}). This was already noted long ago by Manton
and Ruback \cite{curved} (see also \cite{bratek}). Those authors, however, did
not produce a consistent solution taking into account the back-reaction of the
Skyrmion on the geometry. In other words, they did not attempt to solve the
Einstein equations (\ref{einstein}) with the stress-energy tensor
(\ref{tmunu}) generated by a Skyrmion $U$ of the form (\ref{standard1}),
(\ref{3-sphere}), (\ref{pions2.25}). Plugging (\ref{pions2.25}) into
(\ref{3-sphere}) and (\ref{standard1}), the only nonvanishing components of
$T_{\mu}^{\nu}$ are found to be
\begin{equation}
T_{t}^{t}=-\frac{3K(\lambda+\rho_{0}^{2})}{2\rho_{0}^{4}},\qquad T_{\gamma
}^{\gamma}=T_{\theta}^{\theta}=T_{\varphi}^{\varphi}=\frac{K(\lambda-\rho
_{0}^{2})}{2\rho_{0}^{4}}. \label{tmunu0}%
\end{equation}
It can be observed that although the solution $U$ explicitly depends on the
angles $\gamma$, $\theta$ and $\varphi$, the energy-momentum tensor does not,
which means that the back reaction should not upset the isometries of the
background geometry (\ref{metric1}). Solving Einstein's equations with the
energy-momentum tensor (\ref{tmunu0}) algebraically fixes the radius of the
three-dimensional sphere and the cosmological constant in terms of the
remaining parameters in the action,
\begin{equation}
\rho_{0}^{2}=\frac{2\lambda\kappa K}{2-\kappa K},\qquad\Lambda=\frac
{3(2-\kappa K)^{2}}{8\lambda\kappa K}. \label{pions2.3}%
\end{equation}
Hence, the metric (\ref{metric1}) together with the static Skyrmion
(\ref{standard1}), (\ref{3-sphere}) and (\ref{pions2.25}) define a
self-consistent solution of the full Einstein-Skyrme system (\ref{system})
provided the conditions (\ref{pions2.3}) are satisfied. Note that this
requires $\lambda$, $(2-\kappa K)$ and $\lambda$ to have the same sign, which
we take tentatively positive. This solution is the self-gravitating
generalization of the Skyrmions in \cite{curved}. It is useful to stress here that the above constraint is only needed if one wants a static solution with $a(t)=1$. On the other hand, all rest of the analysis of the present paper will hold for generic values of the coupling constants and cosmological constant.

Our result can also be seen as a generalization of the hedgehog ansatz
discussed in \cite{canfora}, that allows for the construction of exact
multi-Skyrmion configurations composed by elementary spherically symmetric
Skyrmions with non-trivial winding number in four-dimensions
\cite{canfora3,canfora4}. 

On any three-dimensional constant time hypersurface, the winding number for
the configuration is
\begin{equation}
w=\frac{-1}{24\pi^{2}}\int\mathrm{Tr}[\epsilon^{ijk}A_{i}A_{j}A_{k}]=+1\,,
\label{winding}%
\end{equation}
which implies that this Skyrmion cannot be continuously deformed to the
trivial $SU(2)$ vacuum, $U=1$ \cite{manton}.


\subsection{Bianchi-IX Self-gravitating Skyrmions}


Remarkably, the above static Skyrmion can be promoted to a time-dependent
solution in which the space-time metric is of the Bianchi type-IXdescribed by
the metric
\begin{equation}
ds^{2}=-dt^{2}+\frac{\rho(t)^{2}}{4}\left[  a^{2}(t)\left(  d\gamma+\cos\theta
d\varphi\right)  ^{2}+d\theta^{2}+\sin^{2}\theta d\varphi^{2}\right]  \ ,
\label{dyn1}%
\end{equation}
where $\rho(t)$ is a global scaling factor and $a(t)$ is a squashing
coefficient. As can be directly verified a Skyrmion of the same form as before
(\ref{standard1}), with $Y^{0}$ and $Y^{i}$ still given by (\ref{3-sphere})
still identically satisfies the Skyrme field equations in a time-dependent
background geometry of the form (\ref{dyn1}). The technical reason why this
happens is that the scale factor $\rho$ and the squashing parameter $a$ depend
only on time, while the Skyrme ansatz depends only on the spatial coordinates.
This is actually consistent with an ansatz for the Skyrmion in which the full
Skyrme system is consistently reduced to a single scalar equation for the
profile \cite{canfora,canfora2}. The Skyrmion in this case still has baryon
charge $+1$.


\section{The time-dependent system}

\label{section3}

The full Einstein-Skyrme field equations (\ref{system}) with the metric
(\ref{dyn1}), reduce to
\begin{subequations}
\label{equd}%
\begin{align}
2a\rho^{2}(2\rho\dot{a}+3a\dot{\rho})\dot{\rho}-2a^{2}\rho^{2}(\Lambda\rho
^{2}+a^{2}-4)-\kappa K[(2\rho^{2}+\lambda)a^{2}+\rho^{2}+2\lambda]  &
=0\ ,\label{equd1}\\
2a^{2}\rho^{2}(2\rho\ddot{\rho}+\dot{\rho}^{2})-2a^{2}\rho^{2}(\Lambda\rho
^{2}+3a^{2}-4)-\kappa K[(2\rho^{2}+\lambda)a^{2}-\rho^{2}-2\lambda]  &
=0\ ,\label{equd2}\\
a\rho^{3}(\rho\ddot{a}+3\dot{\rho}\dot{a})+(a^{2}-1)[\kappa K(\lambda+\rho
^{2})+4a^{2}\rho^{2}]  &  =0\ . \label{equd3}%
\end{align}

The function $a(t)$ describes the deviations from spherical symmetry. For
$a(t)=\pm1$ the spatial sections are three-spheres and so the solution has
full spherical symmetry (which is expected for a gravitating soliton of charge
1 which, on a flat background, has spherical symmetry). Thus, an interesting
question would be whether or not the solutions of the above system of
equations have the property that
\end{subequations}
\begin{equation}
\underset{t\rightarrow+\infty}{\lim}a(t)=\pm1\ , \label{isotr}%
\end{equation}
which would mean that the solutions approach the \textquotedblleft most
symmetric configuration". Alternatively, when this condition is violated
spherical symmetry is \textquotedblleft spontaneously" broken. The flat
Skyrmion of charge $\pm1$ in flat spacetime is isotropic (see, for instance,
\cite{manton}), whereas if Eq. (\ref{isotr})\ does not hold, the gravitating
Skyrmion is not spherically symmetric.

As seen in \cite{canfora6}, assuming $a(t)=\pm1$ turns (\ref{equd1}),
(\ref{equd2}) and (\ref{equd3}) into a consistent one-dimensional dynamical
system for $\rho(t)$, which can be solved explicitly, as discussed in the
following sections. A preliminary analysis of the interesting properties of
this system for generic $a(t)$ was presented in \cite{sergei}. In the present
paper, we will generalize the analysis of \cite{sergei} clarifying the issue
of the final state of the dynamical system. In particular, we address the
question of whether (\ref{isotr}) holds and in which sense this is a stable
condition. The integrability properties of the reduced dynamical system
for$a(t)=\pm1$\ will also be analyzed.


\subsection{Minisuperspace Lagrangian and Hamiltonian}


It is convenient to write the dynamical system made of Eqs. (\ref{equd1}),
(\ref{equd2}) and (\ref{equd3}) using Hamiltonian formalism. The first step is
to observe that Eqs. (\ref{equd2},\ref{equd3}) follow from the variational
principle of the following Lagrange function,
\begin{equation}
L\left(  x^{k},\dot{x}^{k}\right)  =L_{GR}+V_{\Lambda}+V_{Sk}, \label{Lan.01}%
\end{equation}
where $L_{GR}$ is the Lagrangian of general relativity (GR) in the
mini-superspace geometries of the form (\ref{dyn1}), i.e.
\begin{equation}
L_{GR}\left(  a,\dot{a},\rho,\dot{\rho}\right)  =\left(  2\rho^{2}\dot{a}%
\dot{\rho}+3a\rho\dot{\rho}^{2}\right)  +\left(  a^{2}-4\right)  a\rho\,.
\label{Lan.02}%
\end{equation}
It can be checked that varying $L$ with respect to $a$ and $\rho$ yields
(\ref{equd2},\ref{equd3}), where $V_{\Lambda}$ and $V_{Sk}$ are the potential
terms which correspond to the cosmological constant and to the Skyrmion
field,
\begin{equation}
V_{\Lambda}\left(  a,\rho\right)  =\Lambda a\rho^{3}~,~V_{Sk}\left(
a,\rho\right)  =\frac{\kappa K\left(  a^{2}\left(  2\rho^{2}+\lambda\right)
+\rho^{2}+2\lambda\right)  }{2a\rho}\,. \label{Lan.03}%
\end{equation}

Since Lagrangian (\ref{Lan.01}) describes an autonomous system invariant under
time translations generated by $\partial_{t}$, Noether's theorem implies
energy conservation, which turns out to be the left hand side of
(\ref{equd1}). The fact that the energy vanishes reflects the fact that in
General Relativity it is constrained to be zero by invariance under time
reparametrizations, $t\rightarrow\tau(t)$. In a generic time choice the metric
(\ref{dyn1}) is
\begin{equation}
ds^{2}=-N^{2}\left(  \tau\right)  d\tau^{2}+\frac{\rho^{2}(\tau)}{4}\left[
a^{2}(\tau)(d\gamma+\cos\theta d\varphi)^{2}+d\theta^{2}+\sin^{2}\theta
d\varphi^{2}\right]  , \label{eq.SS1}%
\end{equation}
where $N\left(  \tau\right)  d\tau=dt$. In this parametrization the Lagrangian
is
\begin{equation}
\bar{L}\left(  N,a,\dot{a},\rho,\dot{\rho}\right)  =\frac{1}{N}\left(
2\rho^{2}\dot{a}\dot{\rho}+3a\rho\dot{\rho}^{2}\right)  -N\left(
a^{2}-4\right)  a\rho+NV_{\Lambda}+NV_{Sk}. \label{Lan.04}%
\end{equation}
Here it is manifest that the only dynamical degrees of freedom of the system
are metric coefficients $\rho$ and $a$ and the Skyrmion does not bring in new
dynamical variables. Then, varying with respect to the variables $N$, $a$ and
$\rho$ yields equations (\ref{equd1}), (\ref{equd2}) and (\ref{equd3}),
respectively. The corresponding Hamiltonian for this sytem is
\begin{equation}
\mathcal{H}\equiv N\left[  \frac{p_{a}p_{\rho}}{2\rho^{2}}-\frac{3a}{4\rho
^{3}}p_{a}^{2}-\left(  a^{2}-4\right)  a\rho-V_{\Lambda}-V_{Sk}\right]  ,
\label{Lan.05}%
\end{equation}
and the Legendre transformation from $a,\rho,N$ to $p_{a},p_{\rho},\pi_{N}$
reads
\begin{equation}
p_{a}=\frac{2\rho^{2}}{N}\dot{\rho}~,~p_{\rho}=\frac{2\rho^{2}}{N}\dot
{a}+\frac{3a}{\rho}\dot{\rho}~,~\pi_{N}=0. \label{Lan.06}%
\end{equation}



\subsection{Isotropic space-time and the Ermakov-Pinney equation}


For the spherically symmetric space-time $a^{2}=1$, \eqref{equd3} is
identically satisfied, while \eqref{equd1} and \eqref{equd2} reduce to the
following system \cite{canfora6}:
\begin{align}
\dot{\rho}^{2}  &  =\frac{\Lambda}{3}\rho^{2}+\frac{\lambda\kappa K}{2\rho
^{2}}+\frac{\kappa K-2}{2}\ ,\label{firstint1}\\
\ddot{\rho}  &  =\frac{\Lambda}{3}\rho-\frac{\lambda\kappa K}{2\rho^{3}}\ .
\label{firstint2}%
\end{align}

As noted before, (\ref{firstint1}) is the vanishing energy constraint, while
(\ref{firstint2}) is a particular case of the well-known Ermakov-Pinney (EP)
equation\footnote{The EP equation has the form $\ddot{u}+\omega^{2}%
u+bu^{-3}=0$ and admits exact solutions $u=F(y_{1},y_{2})$ where $y_{1},y_{2}$
are the independent solutions of the associated problem $\ddot{y}+\omega
^{2}y=0$ \cite{ErmakovA}.} \cite{Ermakov,Pinney}, which is also found in
various physical systems (see for instance \cite{ErmakovA,ErmakovB}). One of
its features is that it is invariant under a larger than expected symmetry,
$SL(2,R)$ in this case. The representation of the symmetry algebra depends on
whether $\Lambda\lesseqgtr0$. Specifically, the generators of the $SL(3,R)$
Lie algebra are: the autonomous symmetry $\Gamma^{1} =\partial_{t}$, and the
two generators $\Gamma^{2}$ and $\Gamma^{3}$ with representations
\begin{equation}
\Gamma_{(\Lambda>0) }^{2}=\frac{2}{\omega}\sinh(\omega t) \partial_{t}
+\cosh(\omega t) \rho\partial_{\rho}, ~\Gamma_{(\Lambda>0)}^{3} =\frac
{2}{\omega}\cosh(\omega t) \partial_{t} +\sinh(\omega t) \rho\partial_{\rho},
\end{equation}
for positive cosmological constant,~where $\omega^{2}:=4|\Lambda|/3$, or
\begin{equation}
\Gamma_{(\Lambda<0)}^{2} =\frac{2}{\omega}\sin(\omega t) \partial_{t}
+\cos(\omega t) \rho\partial_{\rho}~, ~\Gamma_{(\Lambda<0)}^{3} =\frac
{2}{\omega}\cos(\omega t) \partial_{t} - \sin(\omega t) \rho\partial_{\rho},
\end{equation}
for negative cosmological constant, while when $\Lambda=0$ the generators take
the simple form
\begin{equation}
\Gamma_{(\Lambda=0)}^{2} = 2t\partial_{t} +\rho\partial_{\rho}, ~\Gamma
_{(\Lambda=0)}^{3} = t^{2} \partial_{t} + t\rho\partial_{\rho}.
\end{equation}

The solution of the EP equation (\ref{firstint2}) can be expressed using a
generic solution of the associated linear equation $\ddot{\rho} =\frac
{\Lambda}{3}\rho$ \cite{Pinney,ErmakovA}, as
\begin{equation}
\omega^{2}\rho^{2}=-(K-2)+(K-2+\rho_{0}^{2}\omega^{2})\cosh(\omega t)\pm
\omega\sqrt{2\rho_{0}^{2}(K-2)+2\kappa K\lambda+\rho_{0}^{4}\omega^{2}}%
\sinh{\omega t}\quad(\Lambda>0), \label{erm1}%
\end{equation}
for $\Lambda>0$, and
\begin{equation}
\omega^{2}\rho^{2}=K-2+(-(K-2)+\rho_{0}^{2}\omega^{2})\cos(\omega t)\pm
\omega\sqrt{2\rho_{0}^{2}(K-2)+2\kappa K\lambda-\rho_{0}^{4}\omega^{2}}%
\sin{\omega t}\quad(\Lambda<0), \label{erm2}%
\end{equation}
for $\Lambda<0$, where $\rho_{0}=\rho(0)$ and the second integration constant
has been eliminated by the constraint equation (\ref{firstint1}).

Furthermore, for $\Lambda=0$ the solution is a power law,
\begin{equation}
\rho^{2}=\rho_{1}\left(  t-t_{0}\right)  ^{2}+\rho_{0} \label{erm3}%
\end{equation}
where $\rho_{0}=\frac{\lambda\kappa K}{2-\kappa K}$ and $\rho_{1}=\frac{\kappa
K-2}{2}$.

We note that the functional form of the exact solutions are related with the
representation of the corresponding admitted $\mathbf{sl}(2,R)$ Lie algebra.
From the exact solutions in which $a^{2} (t) =1$ we observe that for positive
cosmological constant the space-time (\ref{dyn1}) has a de Sitter evolution,
while for negative cosmological constant the scale factor $\rho(t)$ is
periodic with frequency $\omega$. Finally for zero cosmological constant and
for $t\rightarrow\infty$ the space-time (\ref{dyn1}) describes the Milne universe.


\subsection{Einstein static universe}


In order to examine the stability properties of the static Einstein universe
around the isotropic solutions (\ref{erm1}) and (\ref{erm2}), let us consider
the critical points for the field equations (\ref{equd1})-(\ref{equd3}). The
critical points of the Hamiltonian (\ref{Lan.05}) are given by the conditions
\begin{equation}
{\frac{\partial V_{eff}}{\partial a}=0~~\text{and~~}\frac{\partial V_{eff}%
}{\partial\rho}=0},~ \label{cp.01}%
\end{equation}
where $V_{eff}=-\left(  a^{2}-4\right)  a\rho-V_{\Lambda}-V_{Sk}$. Taking into
account the additional the constraint (\ref{equd1}) --which reduces to
$V_{eff}=0$, the critical points in the $(\rho,a)$-plane are identified
as\footnote{For $\kappa K<0$ and $\lambda<0$ there would be an additional
possible critical point with $a_{0}\neq0$ at, $\tilde{P}_{0}:\tilde{\rho}%
_{c}=\left[  \frac{8-\kappa K}{2\Lambda}\right]  ^{1/2}=\left[  -\lambda
\frac{a_{0}^{2}+4}{a_{0}^{2}+2}\right]  ^{1/2}~$with~$a_{c}=a_{0}\neq0$ and
$\kappa K=-2a_{0}^{2}(a_{0}^{2}+4)$. The critical point $\tilde{P}_{0}$ can be
neglected in the standard situations where $\kappa K\geq0$.}
\begin{equation}
P_{\pm}:\rho_{c}=\left[  \frac{3(2-\kappa K)}{4\Lambda}\right]  ^{1/2}=\left[
\frac{3\lambda\kappa K\Lambda}{2}\right]  ^{1/4}~,~a_{c}=\pm1. \label{cp.03}%
\end{equation}

Observe that for $\kappa K>2$ the critical points $P_{\pm}$ exist provided
both $\Lambda$ and $\lambda$ are negative, while the opposite happens if
$\kappa K<2$ ($\lambda>0$, $\Lambda>0$). Last but not least, for zero
cosmological constant $P_{\pm}$ exist if and only if $kK=2$ and $\lambda= 0
=\Lambda$.

Finally, we note that these critical points in momentum space are located at
$(p_{a},p_{\rho})=(0,0)$ and therefore they correspond to static
configurations. It should be noted that the critical points $P_{\pm}$ are
exact solutions of the field equations and describe isotropic Einstein static
spacetimes~\cite{Eins1,Eins2} and therefore perturbing around them is a
meaningful test for the stability of the solutions. In the next section we
examine the stability of the critical points $P_{\pm}$ in the linearized
approximation of the time-dependent field equations.


\section{Stability of the spherically symmetric Skyrmion}

\label{isotro}

Let us now study the evolution of an infinitesimal perturbation around the
classical solution near the critical point for $a=1$,
\begin{equation}
a:=1+u(t),\quad\rho:=\rho_{E}+v(t)\,, \label{eq:lin1}%
\end{equation}
where $\rho_{E}$ stands for the exact solution of the EP equation
(\ref{firstint2}), and $u$ and $v$ are the small perturbations. Substituting
this into \eqref{equd} and keeping up to first order in $u$ and $v$, one finds
(from now on we drop the label $E$ from the exact solution $\rho_{E}$)
\begin{subequations}
\label{equdlin}%
\begin{align}
0=  &  \ddot{u}+3\frac{\dot{\rho}}{\rho}\dot{u}+2\left[  K\kappa
+4+K\kappa\lambda\rho^{-2}\right]  \rho^{-2}~u\,,\label{eq-u}\\
0=  &  \ddot{v}+\frac{\dot{\rho}}{\rho}\dot{v}+\rho^{-2}[4+4(\dot{\rho}%
)^{2}-4\kappa K+12\ddot{\rho}\rho-8\Lambda\rho^{2}]~v\nonumber\\
&  +\frac{1}{2}\rho^{-3}[\kappa K-2\Lambda\rho^{4}-\kappa K\lambda
-4\rho+2(\dot{\rho}\rho)^{2}-2\kappa K\rho^{2}+4\rho^{3}\ddot{\rho
}]~u+\nonumber\\
&  +\frac{1}{2}\left[  1+(\dot{\rho})^{2}-\kappa K+2\ddot{\rho}\rho
-\Lambda\rho^{2}\right]  \rho^{-1}+\frac{\kappa K}{4}(-\lambda+2\Lambda
+1)\rho^{-3}\,,\label{eq-v}\\
0=  &  \frac{\dot{\rho}}{\rho}~\dot{u}+\left[  \frac{2}{\rho^{2}}+3\left(
\frac{\dot{\rho}}{\rho}\right)  ^{2}-\frac{\kappa K\lambda}{2\rho^{4}}%
-\frac{\kappa K}{\rho^{2}}-\Lambda\right]  u\nonumber\\
&  +3\frac{\dot{\rho}}{\rho}\frac{\dot{v}}{\rho}+\left[  \frac{3}{\rho^{2}%
}+3\left(  \frac{\dot{\rho}}{\rho}\right)  ^{2}-\frac{3\kappa K}{\rho^{2}}
-2\Lambda\right]  \frac{v}{\rho}+\nonumber\\
&  +\left[  \frac{3}{2\rho^{2}}+3\left(  \frac{\dot{\rho}}{\rho}\right)
^{2}-\frac{3\kappa K}{4\rho^{2}}-\frac{\Lambda}{2}\right]  -\frac{\kappa
K}{4\rho^{4}}(2\Lambda+\lambda)\,. \label{eq-uv}%
\end{align}
Since the solution $\rho(t)$ for (\ref{firstint2}) is explicitly known, Eq.
(\ref{eq-u}) is an ODE for $u(t)$ that can be directly solved. If $\Lambda>0$,
(\ref{erm1}) implies $\rho\sim\rho_{0}e^{(\omega/2)t}$ for $t\rightarrow
\infty$. In this limit, Eq. (\ref{eq-u}) reduces to $\ddot{u}+(3\omega
/2)\dot{u}=0$, whose solution is
\end{subequations}
\begin{equation}
u(t)=u_{0}e^{-(3\omega/2)t}+c, \label{asympt-sol}%
\end{equation}
where $u_{0}$ and $c$ are arbitrary constants fixed by the initial conditions
of the preturbations. This means that for $t\rightarrow\infty$, $a$ can
approach any constant value $1+c$ and there is nothing special about $a=1$ or
$a\neq1$. In fact, Eq. (\ref{eq-u}) has the form of a damped oscillator driven
by an effective harmonic potential $u^{2}[K\kappa+4+K\kappa\lambda\rho
^{-2}]\rho^{-2}$, which vanishes exponentially for $t\rightarrow\infty$, as
well as all of its derivatives. This is a case of the so-called ``neutral
equilibrium" \cite{arnold}.

Having found $u$, Eq. (\ref{eq-v}) can now be solved for $v$. Substituting the
asymptotic expression for $\rho$, (\ref{eq-v}) takes the form
\begin{equation}
0=\ddot{v}+\frac{\omega}{2}\dot{v}-2\omega^{2}v, \label{eqv(t)}%
\end{equation}
whose solution is
\begin{equation}
v(t)=v_{0}e^{mt}%
\end{equation}
with $m=(-1\pm\sqrt{33})\omega/4$. This means that $v(t)$ either vanishes or
blows up for large $t$. Which of the two branches actually occurs is decided
by the constraint equation (\ref{eq-uv}). This last equations is identically
satisfied by the exponentially decaying perturbation and is grossly violated
by the unstable branch. It is therefore verified that under a small
pertrubation around the critical point $\{\rho=\rho_{E},a=1\}$ the solution
settles to $\{\rho=\rho_{E},a=1+c\}$.\newline

\begin{figure}[ptb]
\centering \includegraphics[scale=0.5]{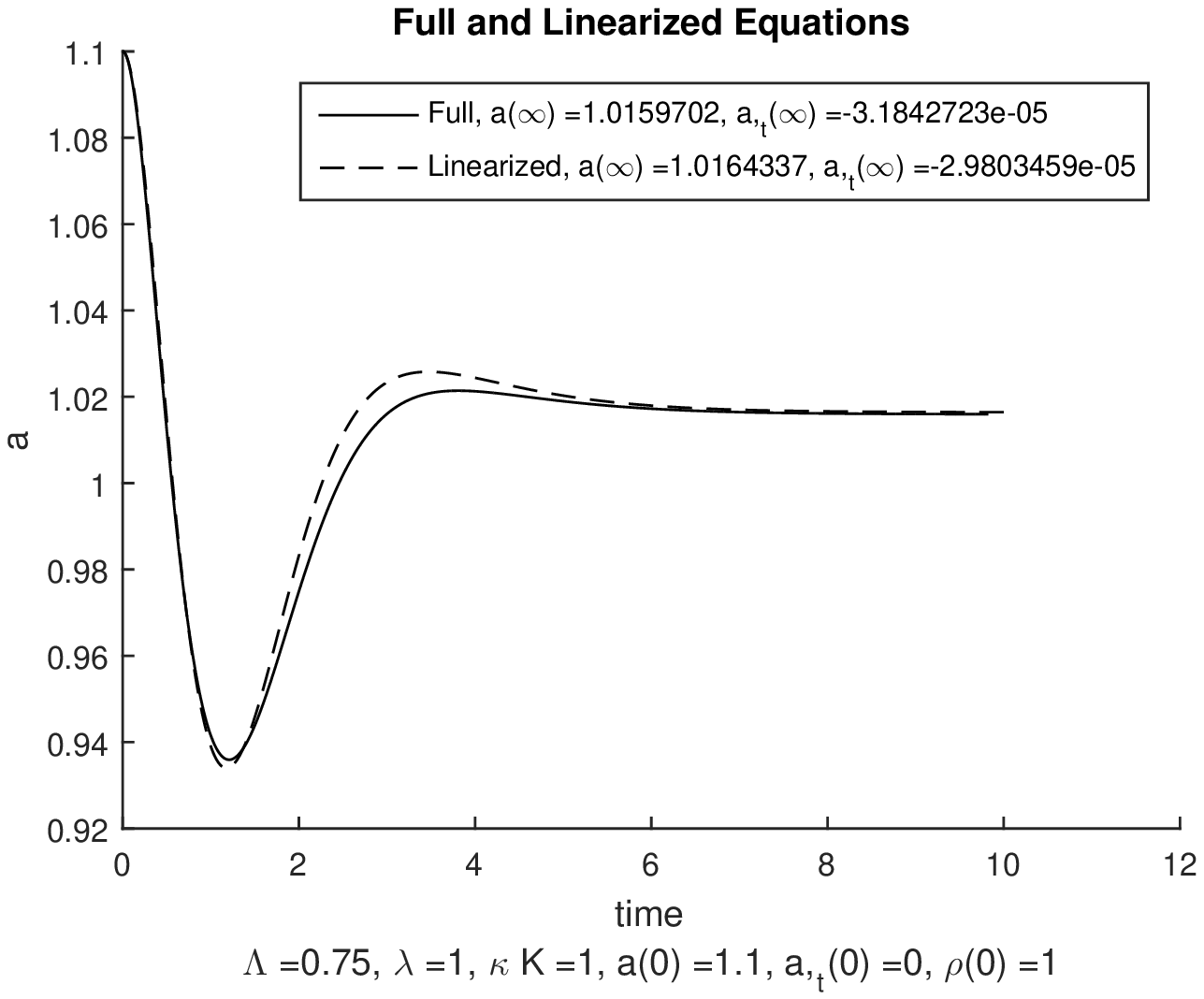}
\centering \includegraphics[scale=0.5]{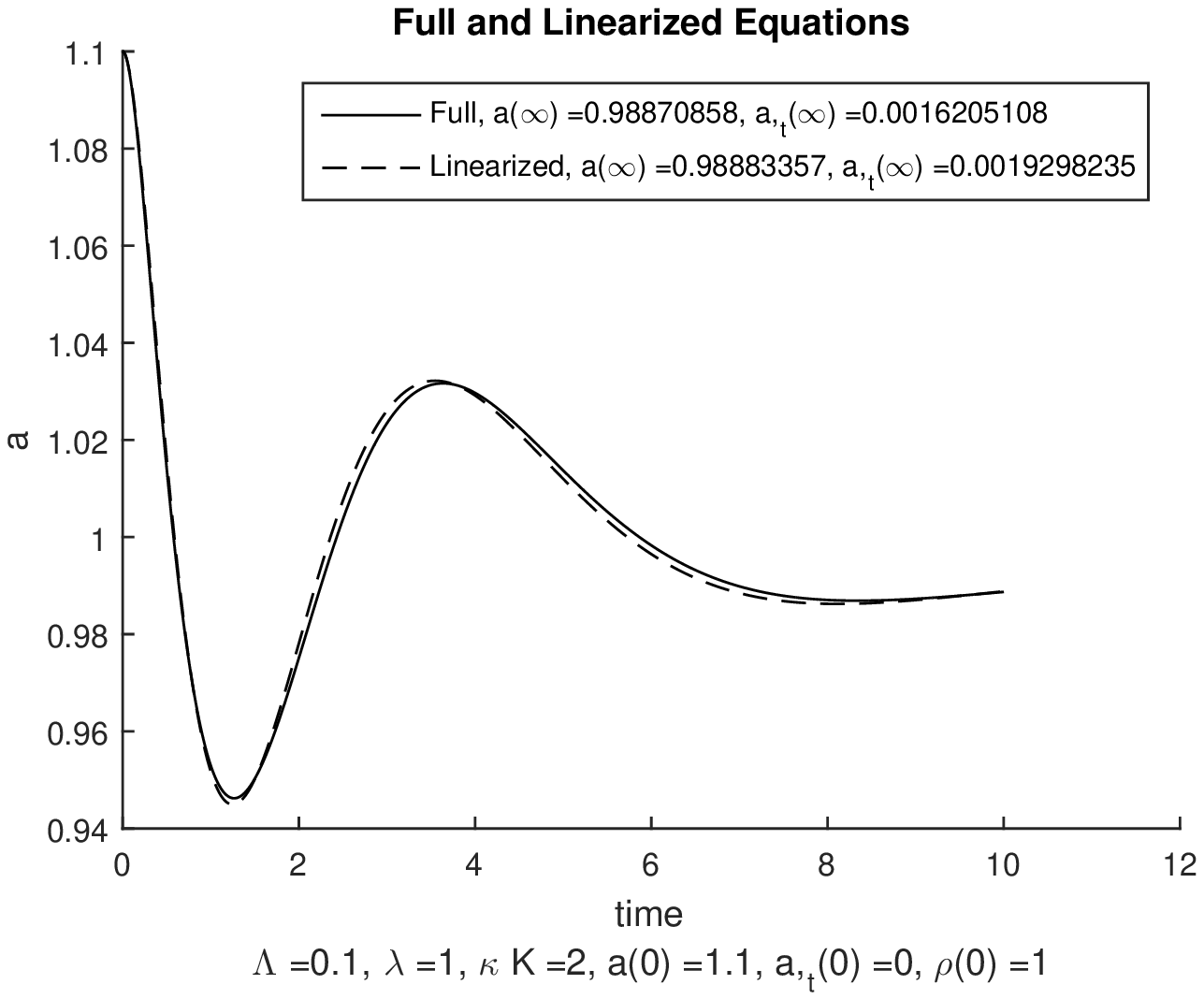}
\centering \includegraphics[scale=0.5]{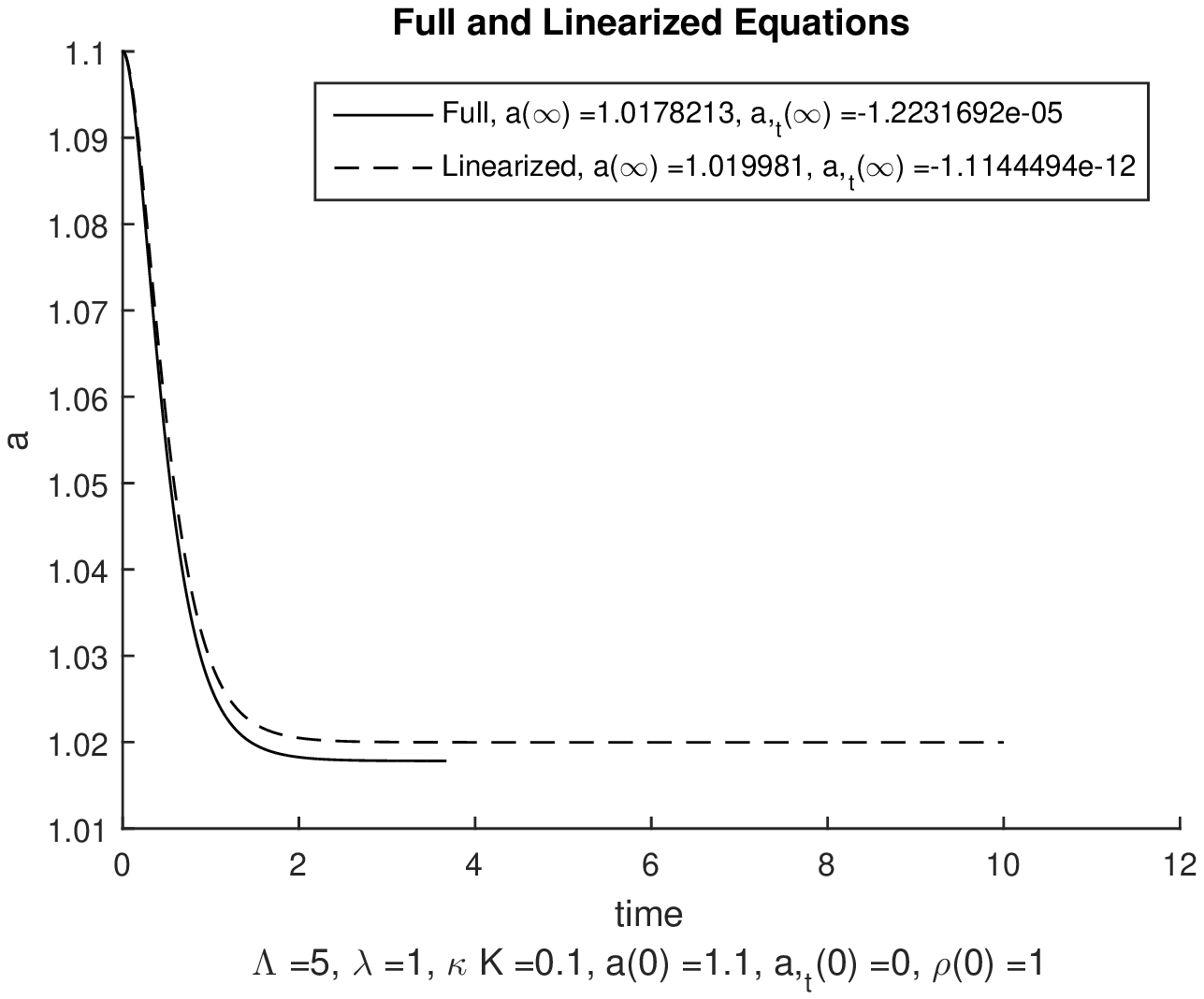}
\centering \includegraphics[scale=0.5]{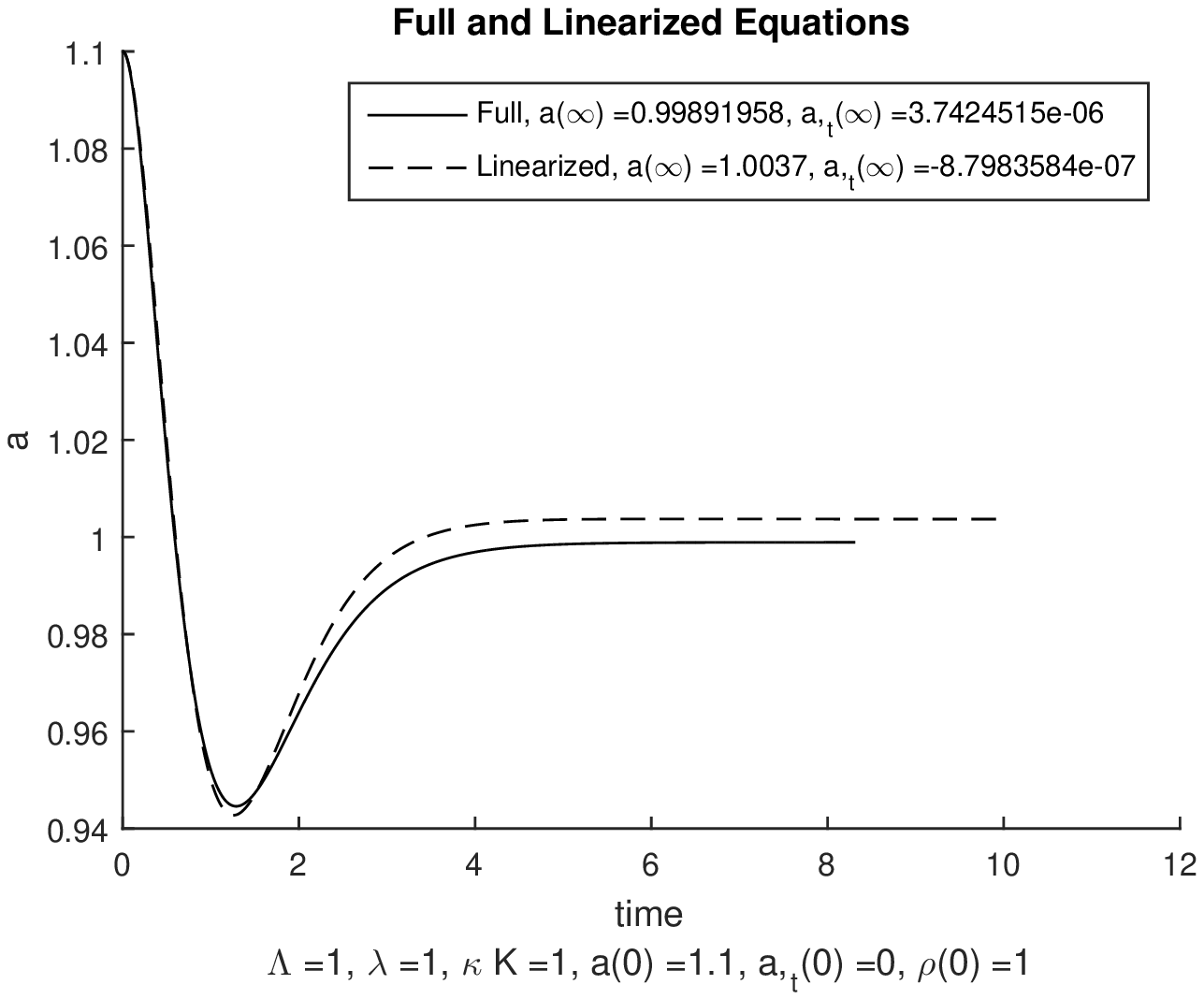} \caption{Qualitative behaviour of the general solution perturbed around the stable solution $a=1$ which is given by the Ermakov-Pinney equation. The plots are for the function $a(t)$ which follow from the total system or the linearized system and for various values of the free parameters where $\Lambda>0$.}%
\label{figure1}%
\end{figure}

\begin{figure}[ptb]
\centering \includegraphics[scale=0.5]{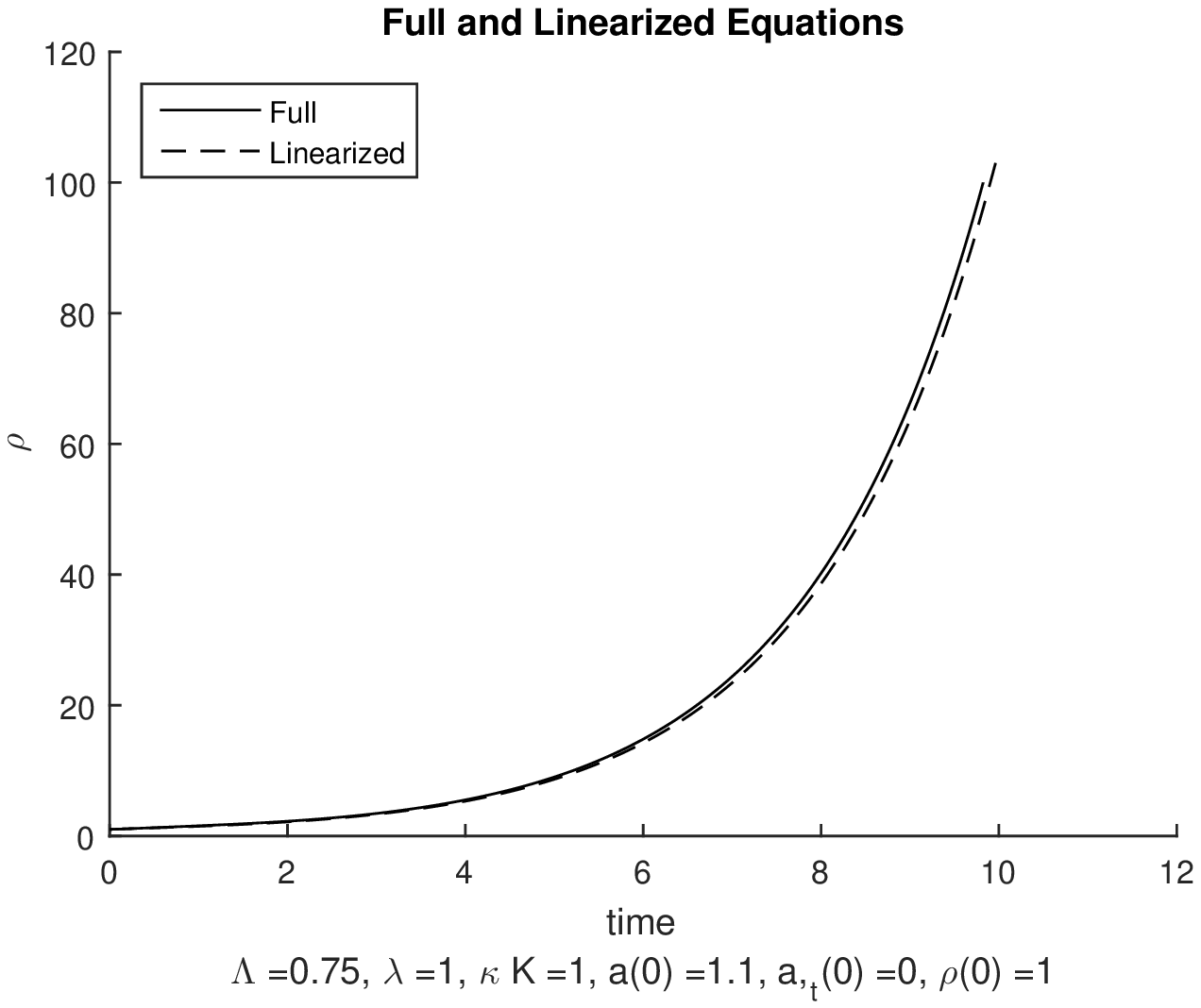}
\centering \includegraphics[scale=0.5]{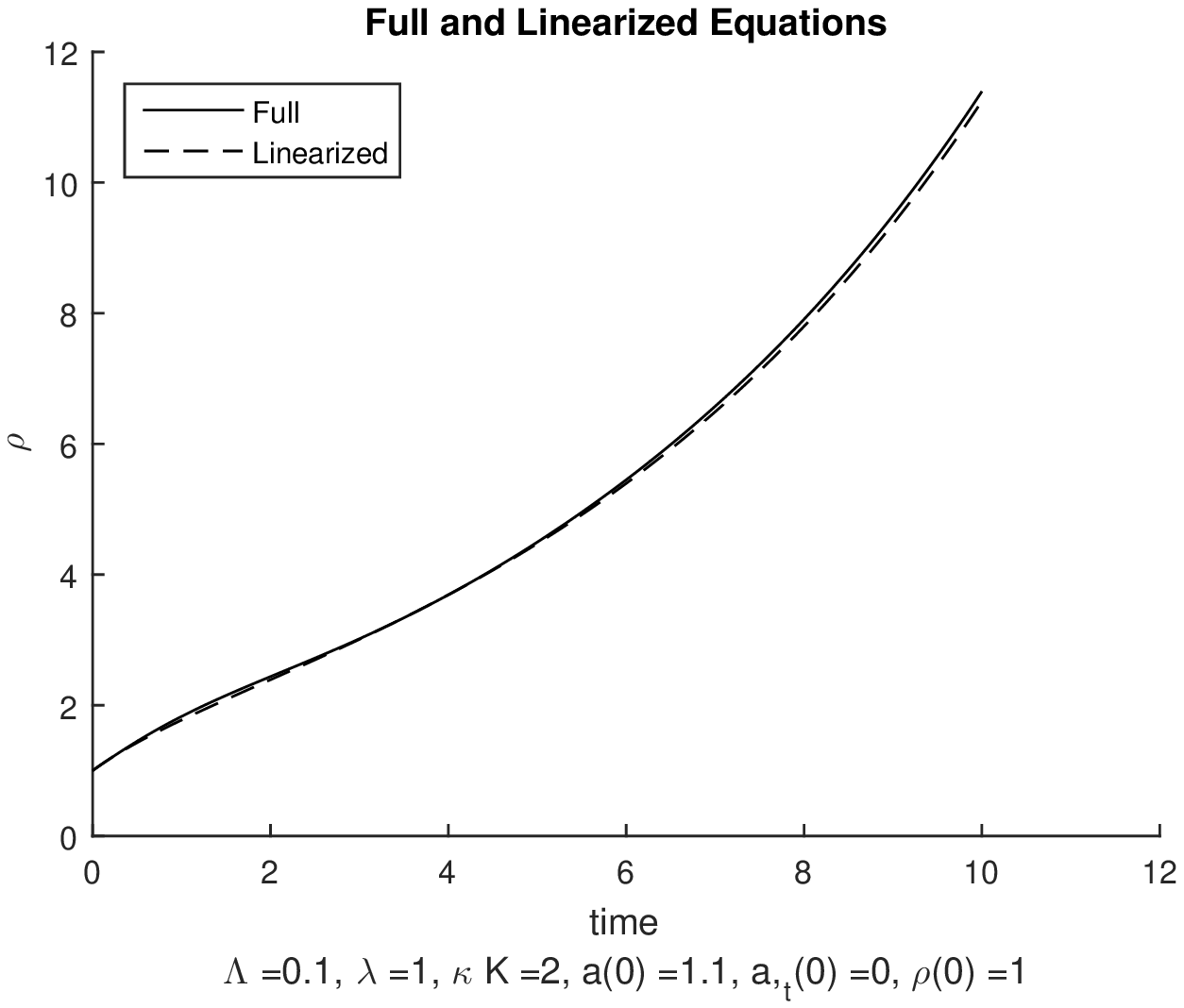} \caption{Qualitative behaviour of the general solution perturbed around the stable solution $a=1$ which is given by the Ermakov-Pinney equation. The plots are for the function $\rho(t)$ which follow from the total system or the linearized system and for various values of the free parameters where $\Lambda>0$.}%
\label{figure2}%
\end{figure}

\begin{figure}[ptb]
\centering \includegraphics[scale=0.5]{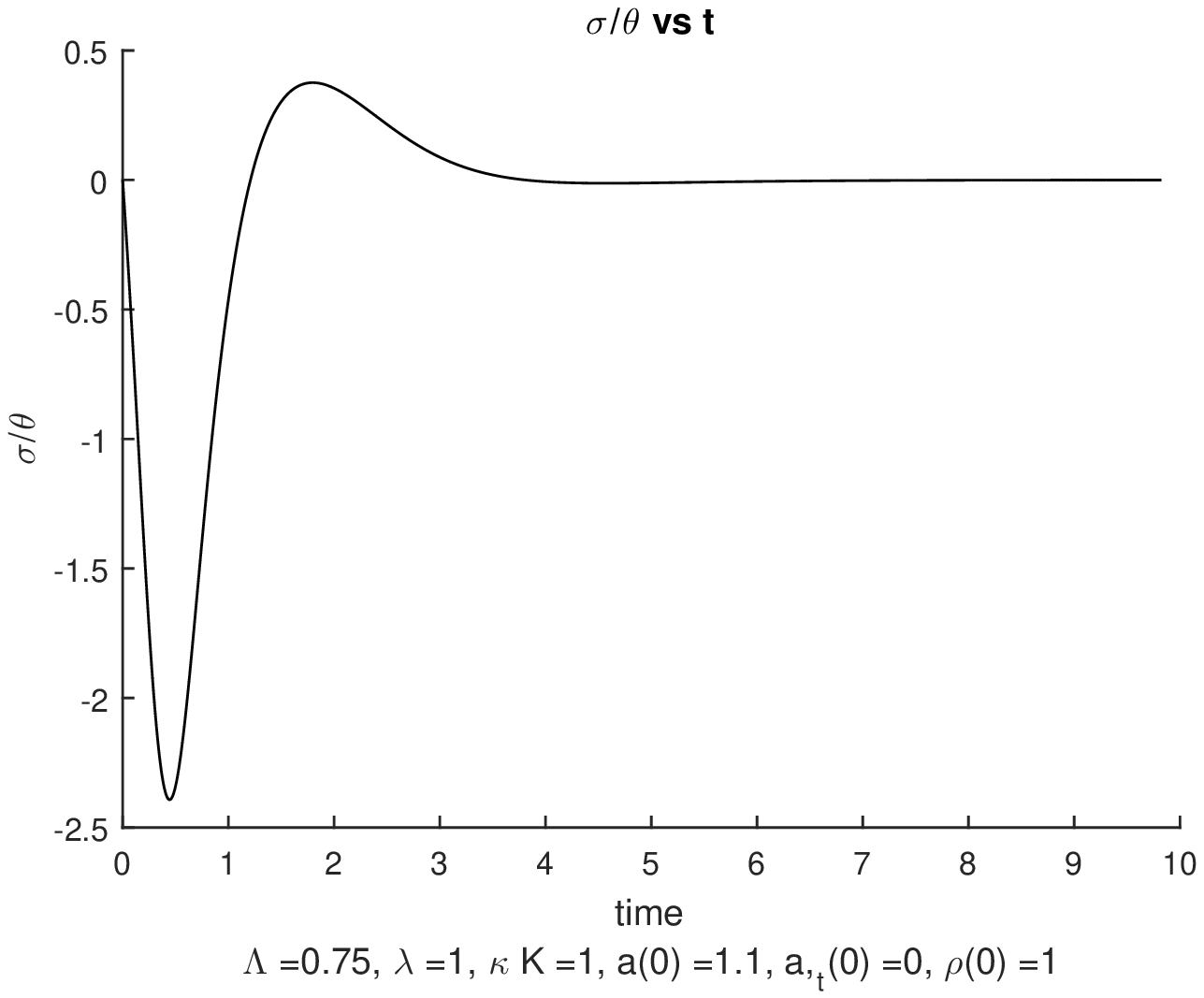}
\centering  \includegraphics[scale=0.5]{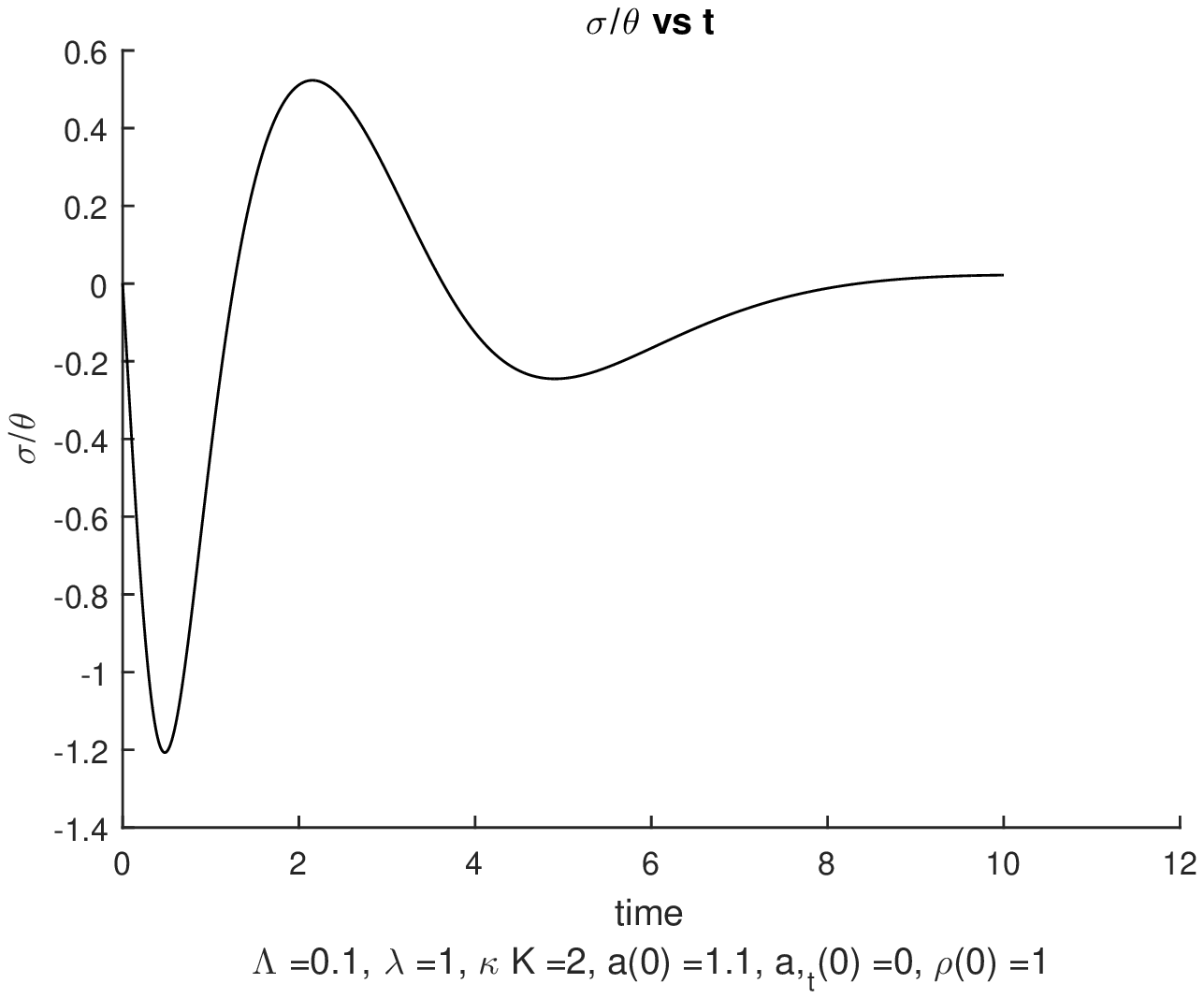} \centering
\includegraphics[scale=0.5]{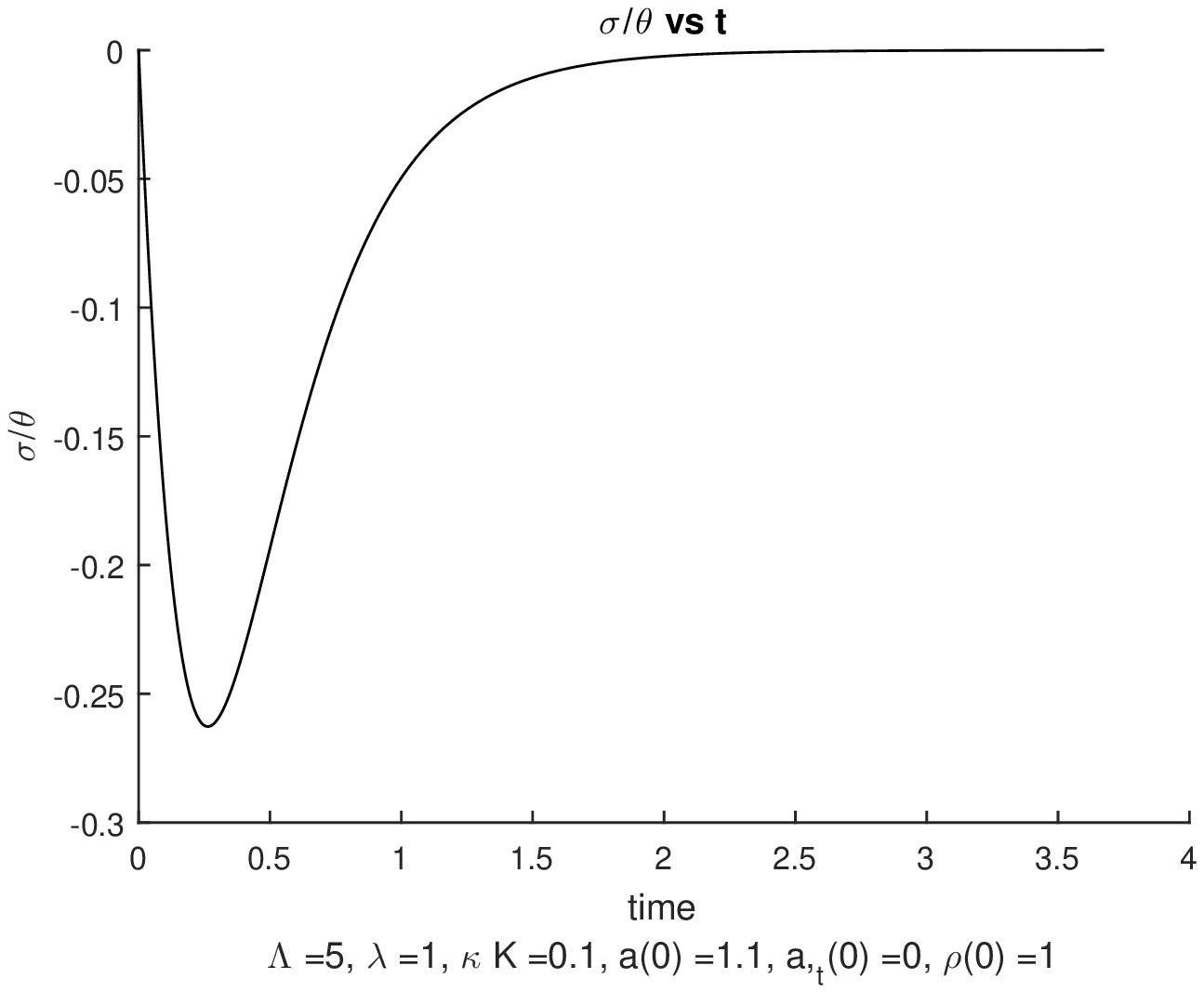}
\centering \includegraphics[scale=0.5]{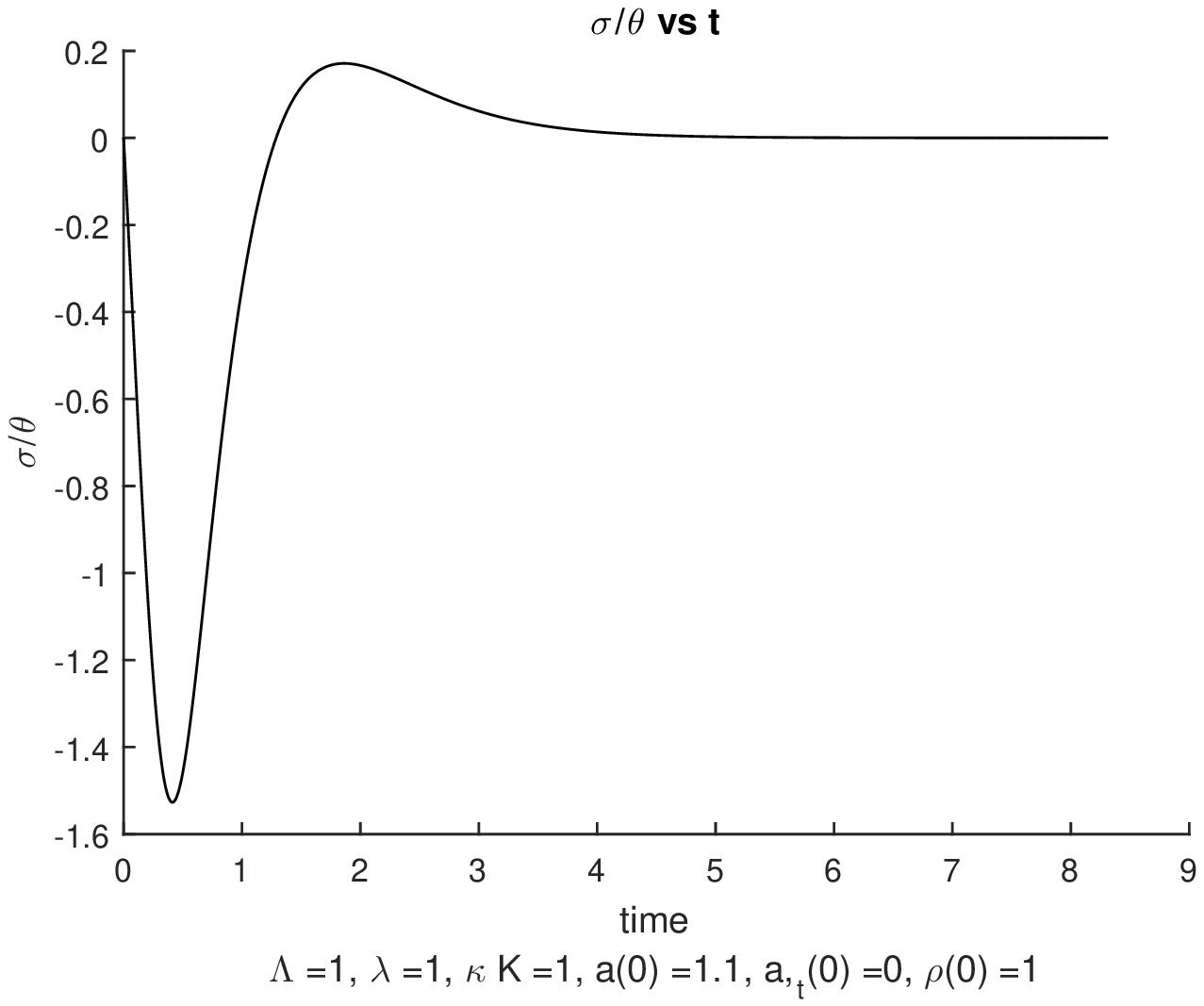} \caption{Qualitative behaviour of the general solution perturbed around the stable solution $a=1$ which is given by the Ermakov-Pinney equation. The plots are for the function $\sigma(t)/\theta(t)$ which follow from the solution of the field equations with various values of the free parameters where $\Lambda>0$. The initial conditions are that of fig \ref{figure1}.}%
\label{figure3}%
\end{figure}

\begin{figure}[ptb]
\centering \includegraphics[scale=0.5]{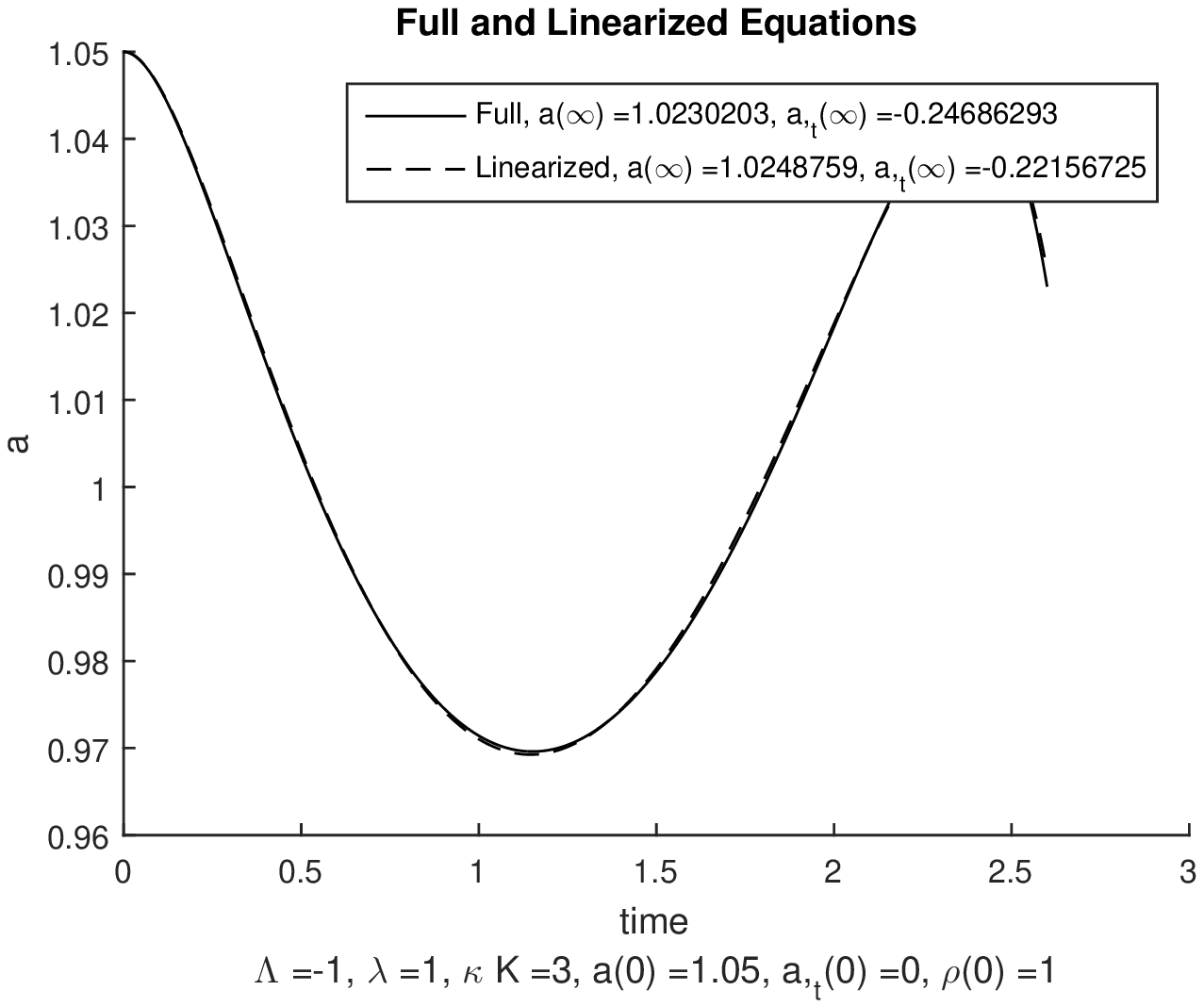}
\centering \includegraphics[scale=0.5]{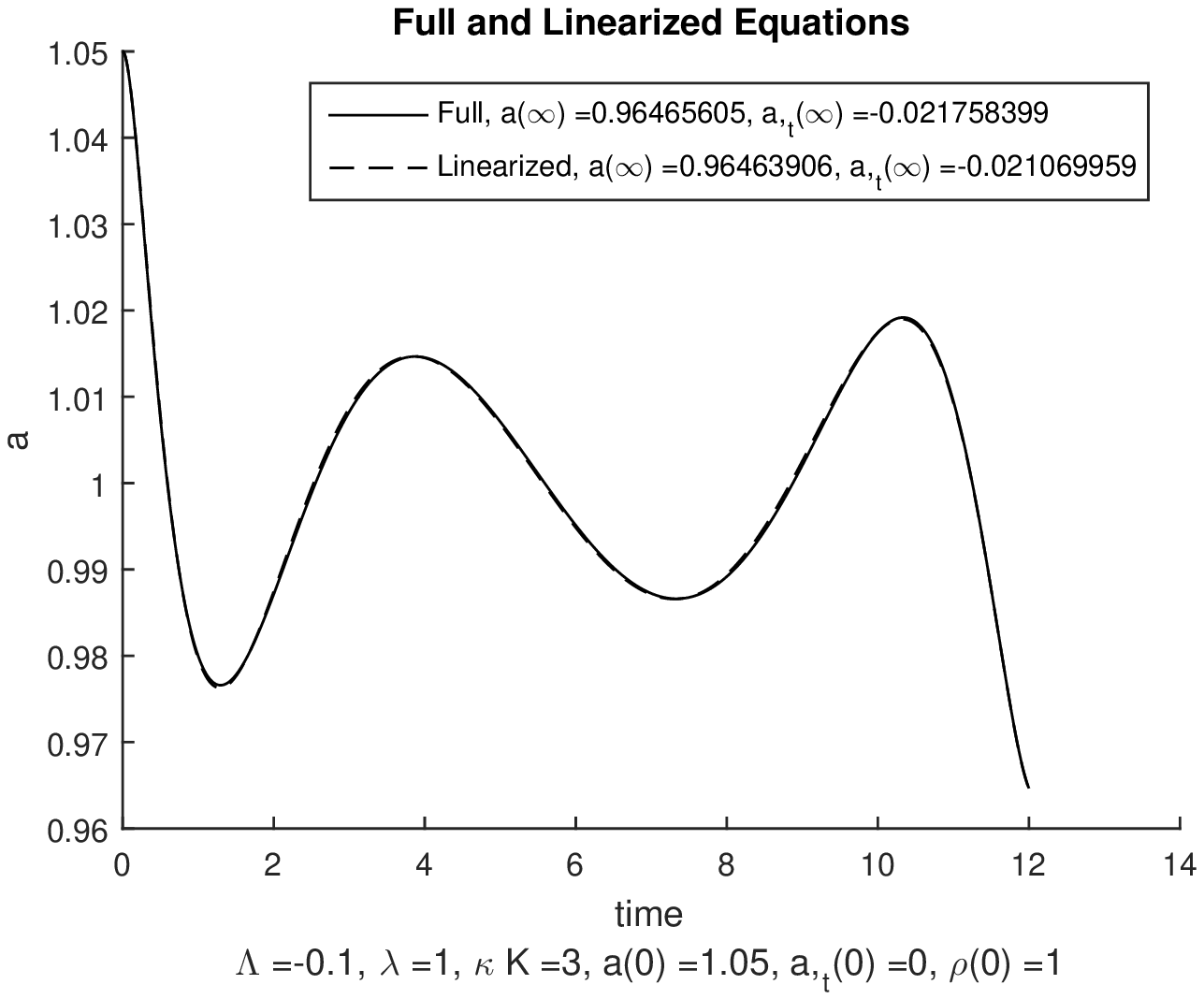} \centering
\includegraphics[scale=0.5]{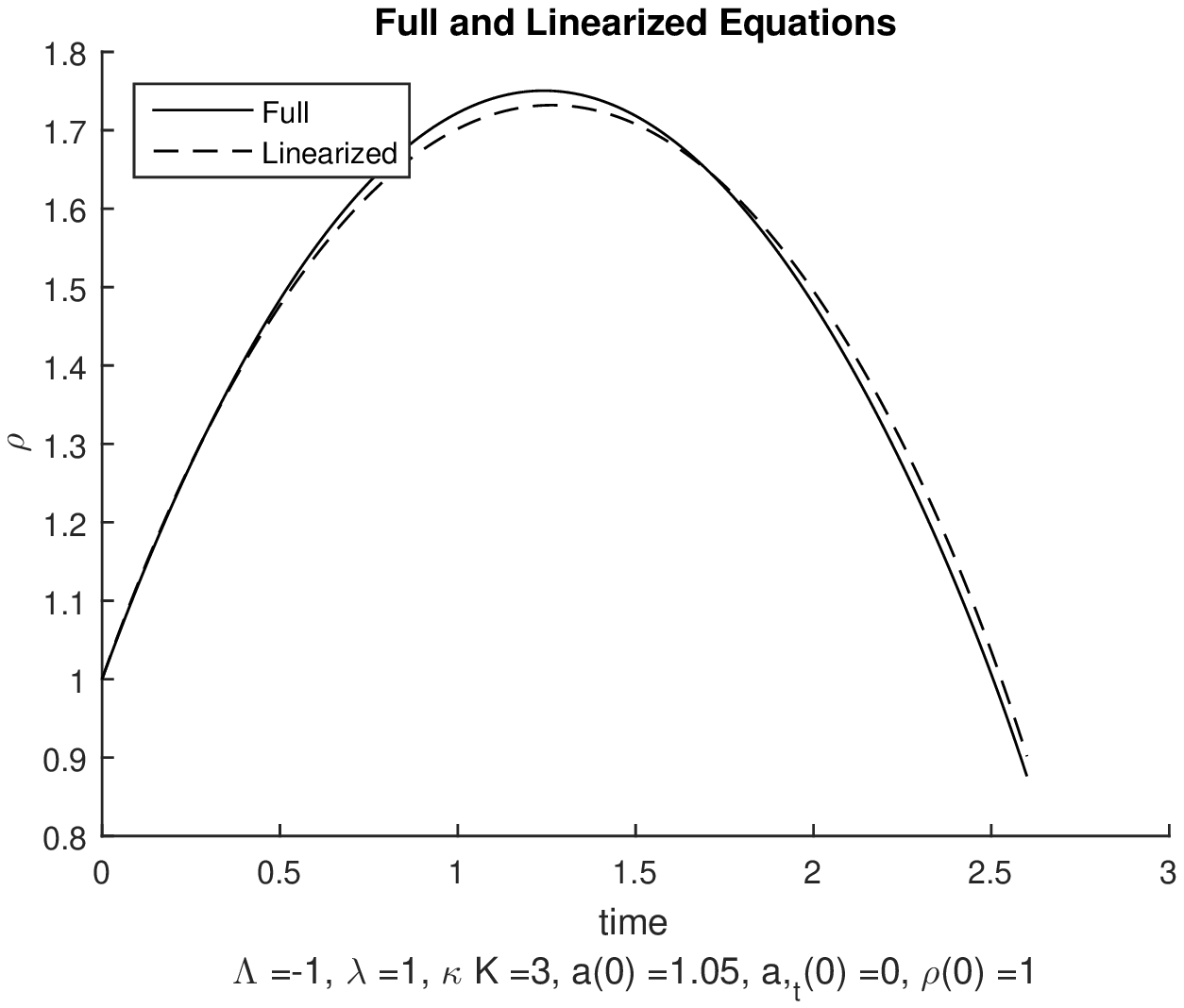}
\centering \includegraphics[scale=0.5]{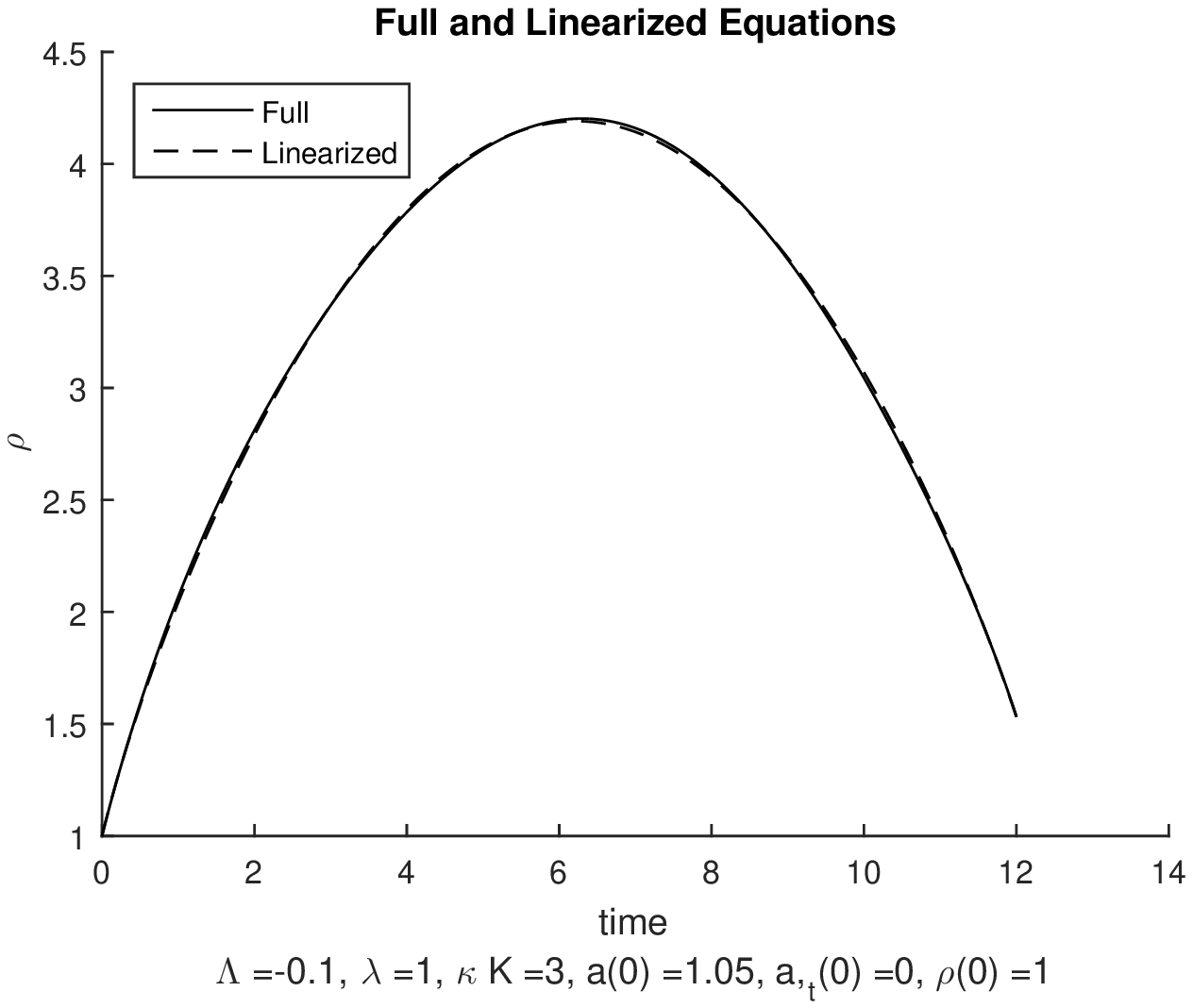}
\centering \includegraphics[scale=0.5]{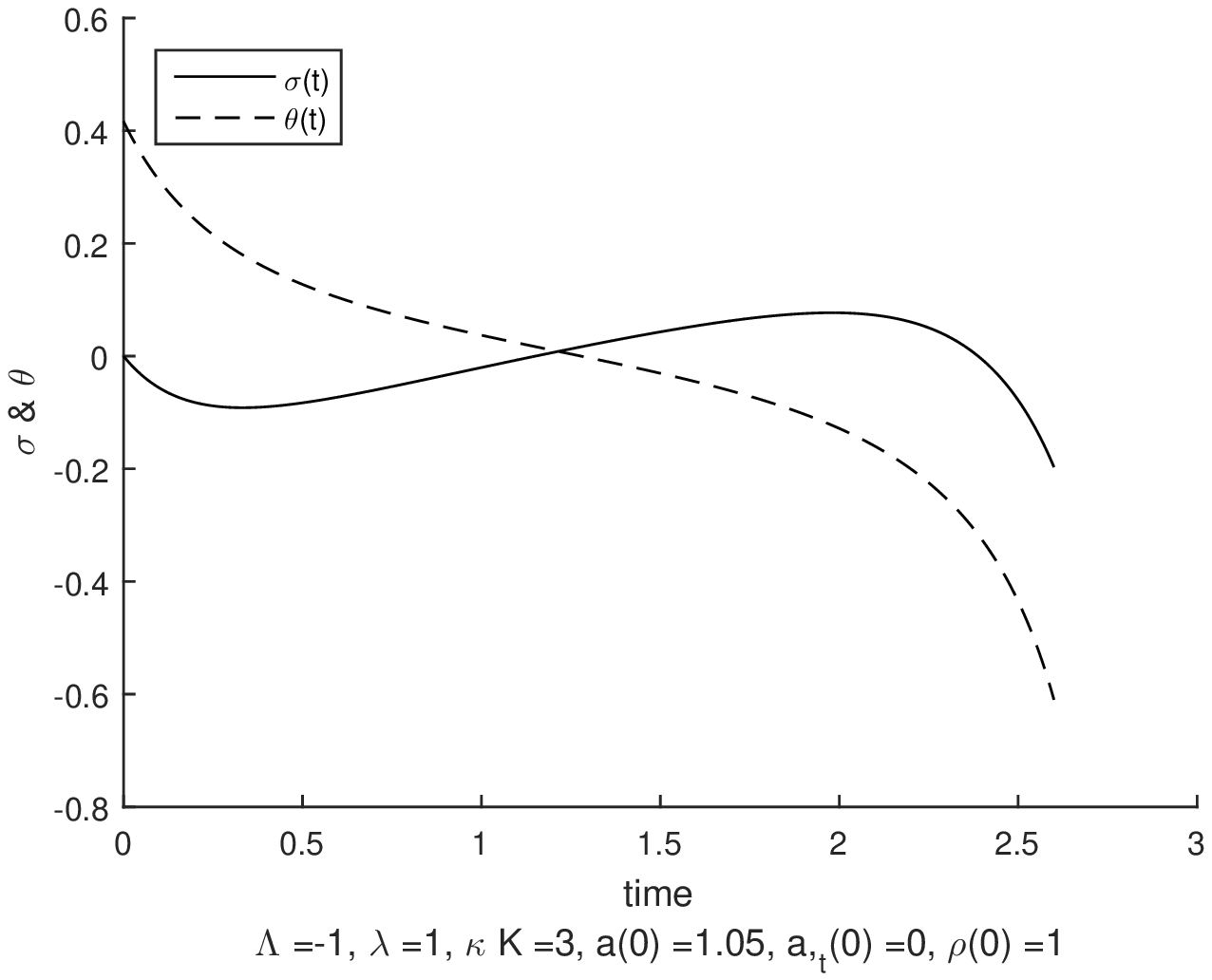}
\centering \includegraphics[scale=0.5]{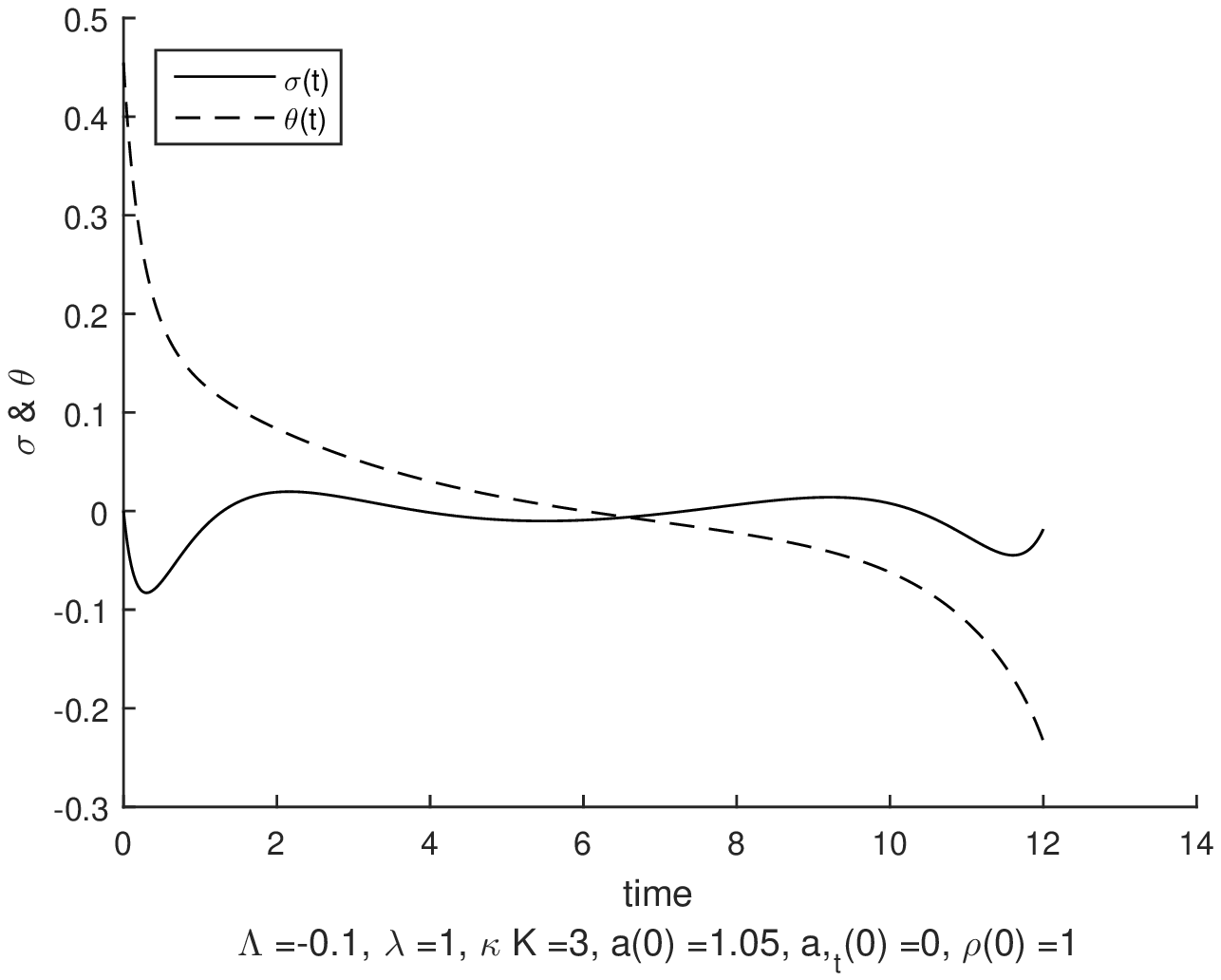} \caption{Qualitative behaviour of the general solution perturbed around the stable solution $a=1$ which is given by the Ermakov-Pinney equation. The plots are for the function $a\left(  t\right)  , \rho\left(  t\right)  $ and $\sigma(t)/\theta(t)$ which follow from the total system or the linearized system and for various values of the free parameters where $\Lambda<0$. }%
\label{figure4}%
\end{figure}

Numerical simulations of the system (\ref{equdlin}) and of the original
equations \eqref{equd} with initial conditions around $a=1$ are summarized in
figures \ref{figure1}-\ref{figure4}. For $\Lambda>0$ figure \ref{figure1}
shows the scalar factor $a(t)$ while \ref{figure2} describes the behavior of
$\rho(t)$. These figures show that for large $t$, $\rho(t) \rightarrow+\infty$
and $a(t) \rightarrow a_{0}=\pm1 +c$, where $c$ is the constant of
(\ref{asympt-sol}) that can take any value depending on the initial
conditions. Although the solution is not strictly stable around $a^{2} (t)
=1$, the space-time for $t\rightarrow\infty$ is an infinitely large squashed
sphere and therefore to a good approximation, locally indistinguishable from a
sphere. The main reason is that, when $\Lambda>0$, the terms in the dynamical
system which lead to the instability of the isotropic solution are suppressed
for $t\rightarrow+\infty$ so that, effectively, such ``destabilizing" terms
only \textit{act for a finite time} after which the value of $a(t)$ becomes
constant as we will see in the next Section. The peculiar neutral equilibrium
feature of the present system means that if the initial data are close to
$a^{2} =1$, for later times $a^{2}(t)$approach $a^{2}_{0}$ in the vicinity of
$1$.

A numerical simulation for the case $\Lambda<0$ is shown in figure
\ref{figure4}. In this case $\rho(t)$ is periodic and may vanish for specific
initial conditions. In that case, the solution $u(t)$ from (\ref{eq-u})
reaches a singularity for which $\ddot{u}(t) \rightarrow\infty$. It is
straightforward to see that in general $u(t)$ is not a decreasing function
which means that the EP solution is unstable.

\subsection{Asymptotically isotropic space-time}


Let us now examine the isotropization of spacetime for large $t$. According to
\cite{Haw}, if a solution of the field equations (\ref{equd1})-(\ref{equd3}),
in the limit $t\rightarrow+\infty$, satisfies the conditions: (a) the global
scale factor $\rho\left(  t\right)  $ is going to infinity, i.e.
$\rho(t)\rightarrow+\infty$, (b) the anisotropic parameter $a\left(  t\right)
$ becomes constant, $a(t)\rightarrow a_{0}$, (c) the weak energy condition is
not violated $T^{00}>0$, while it holds $T^{0i}/T^{00}\rightarrow0$ and (d)
the ratio of the shear $\sigma$ with the expansion rate $\theta$ vanishes,
i.e. $\frac{\sigma}{\theta}\rightarrow0$, then the space-time (\ref{dyn1})
will be asymptotically isotropic. $T^{\mu\nu}~$is the energy momentum tensor,
the kinematic quantities $\sigma~$and $\theta$ are defined by the observer
$u^{\mu}=\delta_{t}^{\mu}~\left(  u^{\nu}u_{\nu}=-1\right)  $, such as
$\sigma^{2}=\sigma_{\mu\nu}\sigma^{\mu\nu}$, where $\sigma_{\mu\nu}=u_{\left(
\kappa;\lambda\right)  }\left(  h_{\mu}^{\kappa}h_{\nu}^{\lambda}-\frac{1}%
{3}\theta h_{\mu\nu}\right)  $ and $\theta=u_{\left(  \mu;\nu\right)  }%
h^{\mu\nu}$ in which $h_{\mu\nu}$ is the projective tensor $h_{\mu\nu}%
=g_{\mu\nu}+u_{\mu}u_{\nu}$.

In figure \ref{figure3} the evolution of the anisotropy parameter
$\sigma/\theta$ is presented from where we can see that the ration vanishes.

\begin{figure}[ptb]
\centering\includegraphics[scale=0.5]{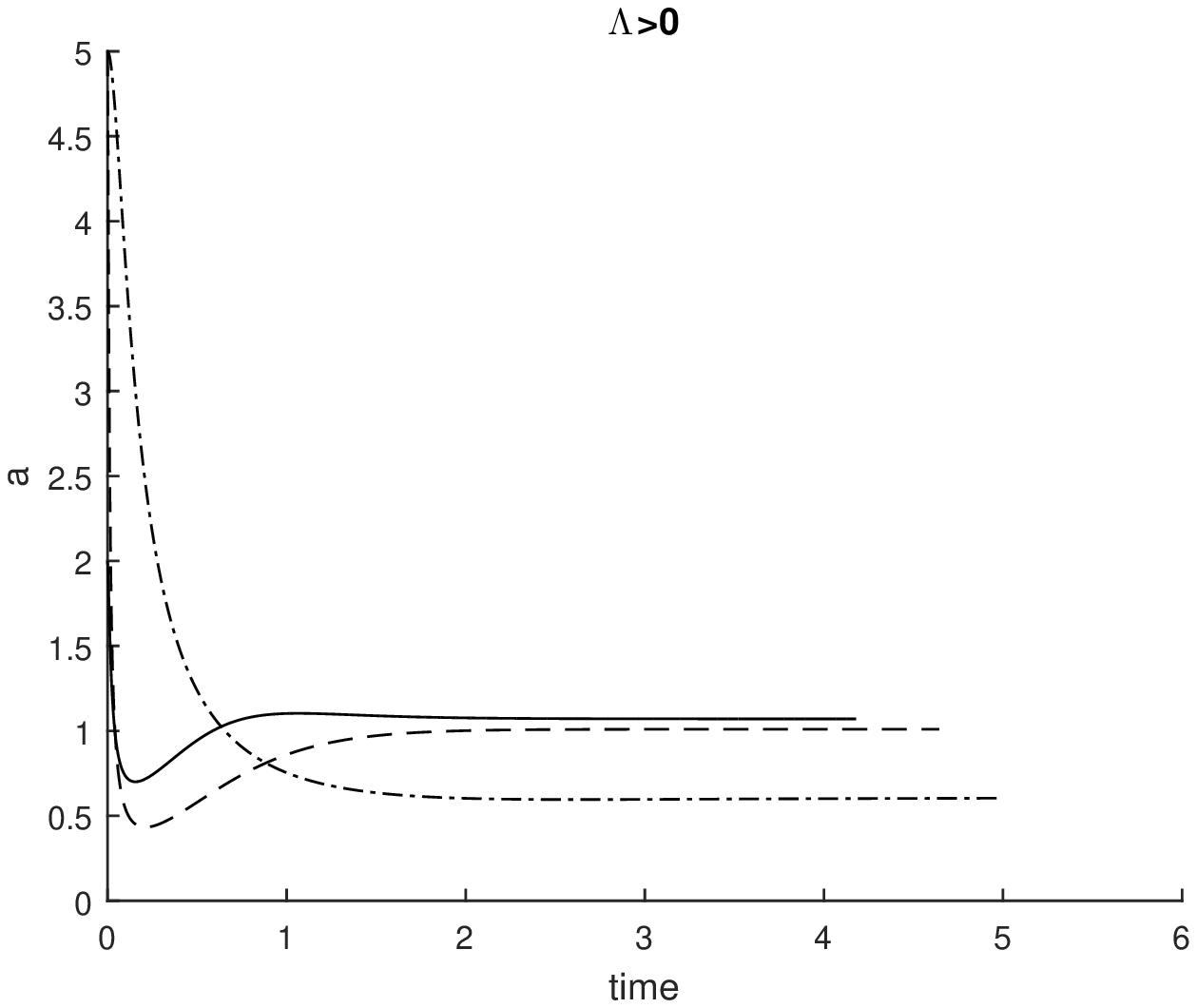}
\centering \includegraphics[scale=0.5]{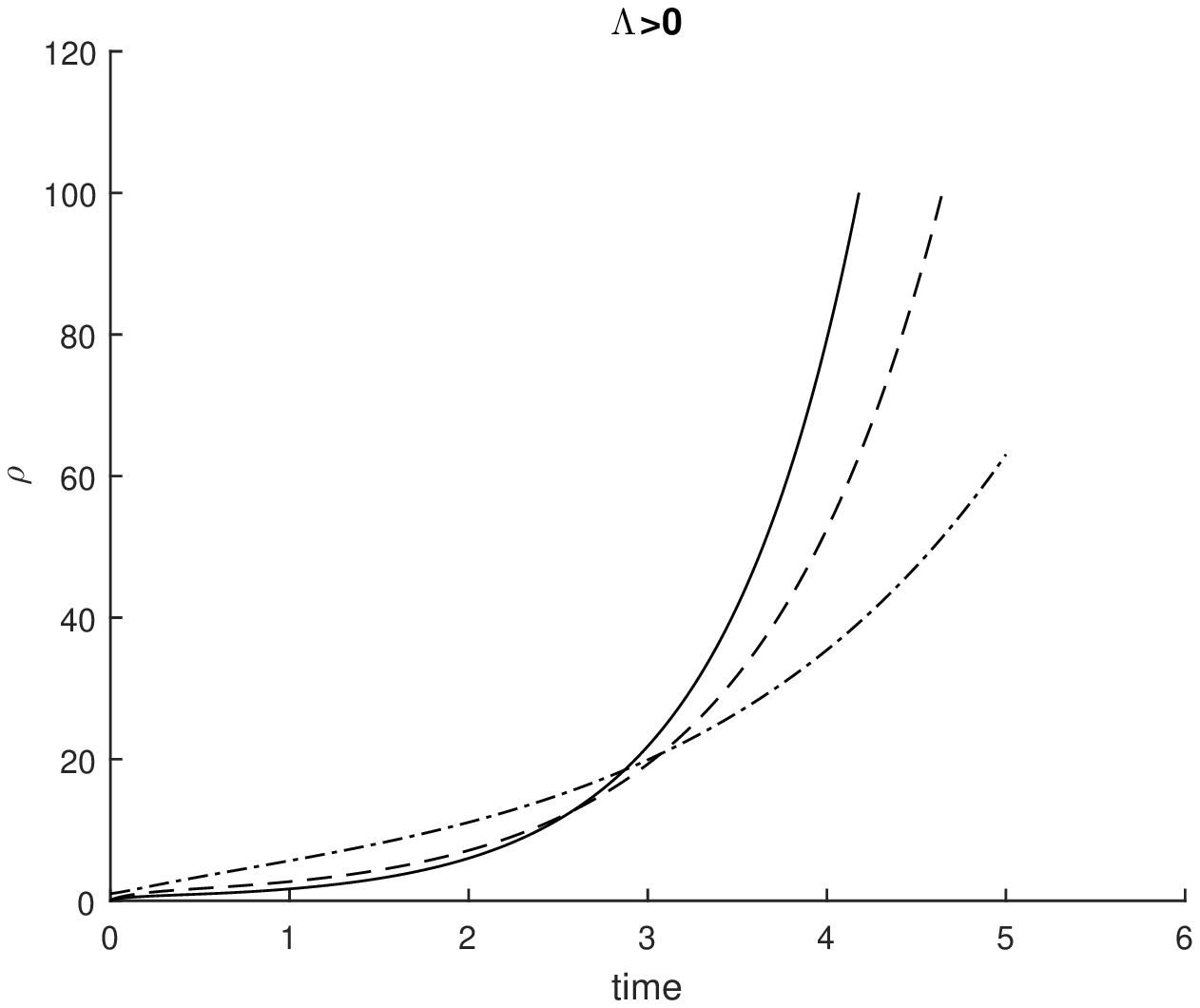}\caption{Qualitative
behaviour of the solution of the field equations (\ref{equd1})-(\ref{equd3})
for initial conditions far for $a\left(  0\right)  =1$. The left plot is for
the scale factor $a\left(  t\right)  $, while the right plot is for the scale
factor $\rho\left(  t\right)  $. The lines are for various values of the free
parameters, where $\Lambda>0$. }%
\label{figure5}%
\end{figure}

\begin{figure}[ptb]
\centering\includegraphics[scale=0.5]{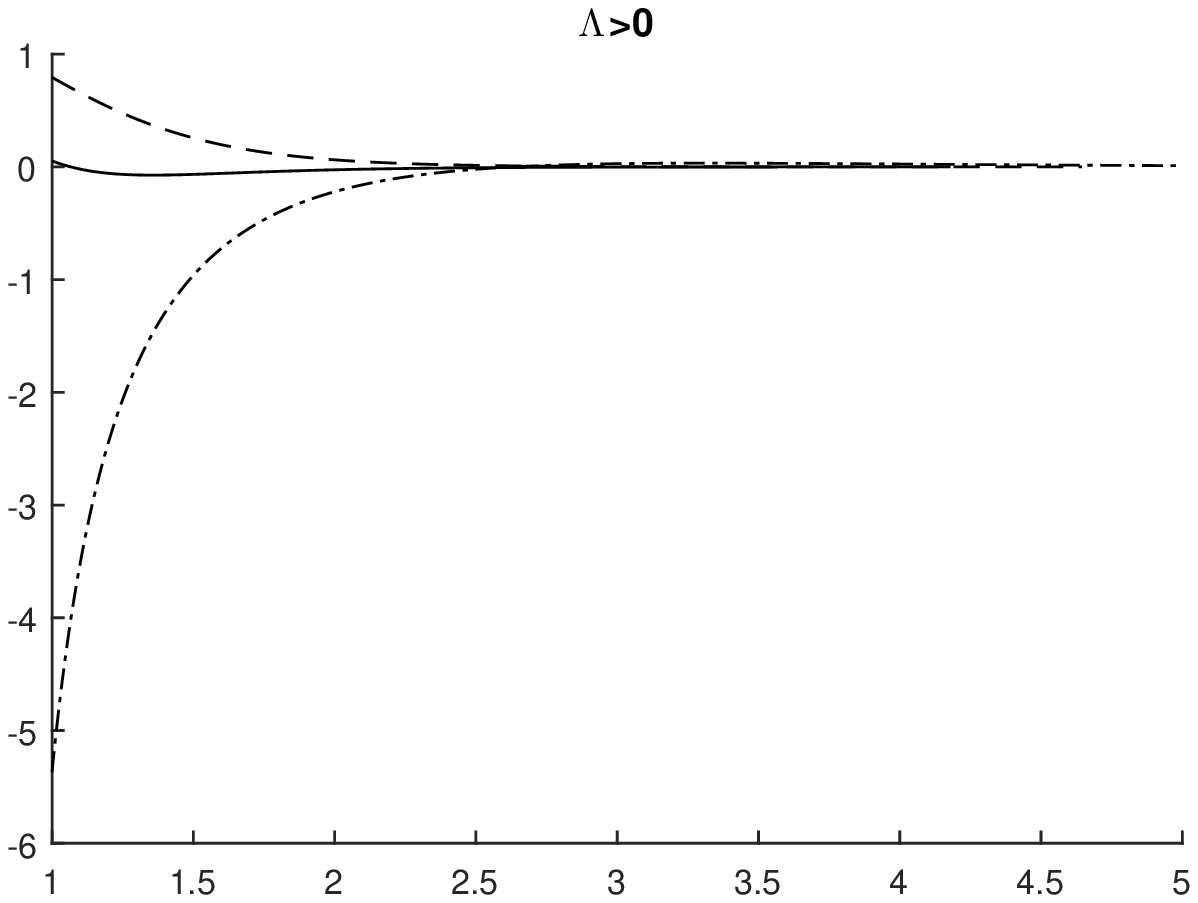}
\caption{Qualitative behaviour of the anisotropic parameter $\sigma\left(
t\right)  /\theta\left(  t\right)  $, for the solutions of fig. \ref{figure5}%
.}%
\label{figure6}%
\end{figure}

For $\Lambda>0$, conditions (a), (b) and (c) are satisfied. Figure
\ref{figure5} shows the evolution of $a(t)$, $\rho(t)$ for the system
(\ref{equd1})-(\ref{equd3}) with $\Lambda>0$ and initial conditions far from
the point $a_{0} =1$. We observe that as $t\rightarrow+\infty$, conditions (a)
and (b) are satisfied, while figure \ref{figure6} shows that condition (d) is
also satisfied, because $\sigma=\frac{\sqrt{6}}{3}\left\vert \frac{\dot{a}}%
{a}\right\vert $, which implies that space-time is asymptotically isotropic.
On the other hand, for $\Lambda<0$, condition $(c),~$ $T^{00}>0,$ can be
violated which means that the ``isotropization" is not guaranteed.

The present analysis shows that in general, the exact solution (\ref{erm1})
with $a(t)=\pm1$ is unstable. However the spacetime is asymptotically
isotropic for large values of $t$. That means that in the late-time the only
fluid-term which survives is that of the cosmological constant. That result
revises the previous analysis of \cite{sergei}.

\section{Conclusions}

\label{conc}

We have analyzed the gravitating, time-dependent analytic solutions of the
Einstein-Skyrme system with topological charge one introduced in
\cite{canfora6}. In particular, we have shown that these solutions --whose
analogues in flat space-times would be spherically symmetric--, reach an
isotropic asymptotic state for $t\rightarrow+\infty$. This question was also
analyzed numerically in \cite{sergei}. In addition, we have shown that the
isotropic solution, given by the Ermakov-Pinney equation, itself is not stable
configuration, but a state of neutral equilibrium, like a spontaneously broken
vacuum. Thus, the isotropy of the charge $1$ Skyrmion on flat spaces may be
broken by the coupling with Einstein gravity. However, despite this fact, the
asymptotic solutions for $\Lambda>0$ of the dynamical system describing the
time-dependent gravitating Skyrmion are asymptotically isotropic in large
scales. The main reason is that, when $\Lambda>0$, the \textquotedblleft
destabilizing" terms in the dynamical system (leading to the instability of
the isotropic solution) are suppressed for $t\rightarrow+\infty$.
Consequently, such terms only act for a finite amount of time after which the
value of $a(t)$ freezes. To the best of the authors' knowledge, this is the
first explicit example of a symmetry breaking induced by the coupling with
Einstein gravity of a topological soliton (which on flat spaces would be
isotropic) in a realistic theory such as the Skyrme model. Moreover, we have
discussed in detail the integrability of the isotropic solution in terms of
the Ermakov-Pinney\ system.

\begin{acknowledgments}
This work has been funded by the Fondecyt grants 1160137, 1121031, 1130423,
1140155, 1141073 and 3140123, together with the CONACyT grants 175993 and
178346. The Centro de Estudios Cient\'{\i}ficos (CECs) is funded by the
Chilean Government through the Centers of Excellence Base Financing Program of
Conicyt. AP acknowledges financial support of FONDECYT grant 3160121, and
Durban University of Technology for hospitality while part of this work was performed.
\end{acknowledgments}


\begin{thebibliography}{99}                                                                                               %


\bibitem {skyrme}T. Skyrme, Proc. R. Soc. London A 260, 127 (1961); Proc. R.
Soc. London \textbf{A 262}, 237 (1961); Nucl. Phys.\ \textbf{31}, 556 (1962).


\bibitem {witten0}E. Witten, Nucl. Phys.\ B \textbf{223}, 422 (1983); Nucl.
Phys. \textbf{B 223}, 433 (1983).

\bibitem {finkrub}D. Finkelstein and J. Rubinstein, J. Math. Phys, \textbf{9},
1762 (1968).

\bibitem {multis2}H. Weigel, Chiral Soliton Models for Baryons, (Springer
Lecture Notes 743)

\bibitem {manton}N.~Manton and P.~Sutcliffe, Topological Solitons, Cambridge
University Press, Cambridge, (2007).


\bibitem {susy}David I. Olive and Peter C. West (Editors), Duality and
Supersymmetric Theories, Cambridge University Press, Cambridge, (1999)

\bibitem {giulini}D. Giulini, Mod. Phys.Lett. \textbf{A 8}, 1917 (1993).

\bibitem {bala2}A.P. Balachandran, H. Gomm and R.D. Sorkin, Nucl. Phys.
\textbf{B 281}, 573 (1987).

\bibitem {bala0}A.P. Balachandran, A. Barducci, F. Lizzi, V.G.J. Rodgers and
A. Stern, Phys. Rev. Lett. 52, 887 (1984).

\bibitem {bala1}A.P. Balachandran, F. Lizzi, V.G.J. Rodgers and A. Stern,
Nucl. Phys. \textbf{B 256}, 525 (1985).

\bibitem {ANW}G. S. Adkins, C. R. Nappi and E. Witten, Nucl. Phys. \textbf{B
228}, 552 (1983).

\bibitem {guada}E. Guadagnini, Nucl. Phys. \textbf{B 236}, 35 (1984).

\bibitem {rec1}Adam, J. Sanchez-Guillen and A. Wereszczynski, Phys. Lett.
\textbf{B 691}, 105 (2010).

\bibitem {rec2}C. Adam, T. Klahn and C. Naya, J. Sanchez-Guillen, R. Vazquez
and A. Wereszczynski, Phys. Rev. \textbf{D 91}, 125037 (2015).

\bibitem {curved1f}D. Auckly and M. Speight, Commun. Math. Phys. \textbf{263,}
173 (2006).

\bibitem {curved2f}S. Krusch and M. Speight, Commun. Math. Phys. \textbf{264,} 391~(2006).

\bibitem {lucock}H. Luckock, I. Moss, Phys. Lett. \textbf{B 176,} 341 (1986).

\bibitem {droz}S. Droz, M. Heusler and N. Straumann, Phys. Lett. \textbf{B
268,} 371 (1991).

\bibitem {bh01}S. Droz, M. Heusler and N. Straumann, Phys. Lett. \textbf{B
268,} 371 (1991)

\bibitem {bh02}M. Heusler, S. Droz and N. Straumann, Phys. Lett. \textbf{B
285,} (1992).

\bibitem {droz2}S. Droz, M. Heusler, N. Straumann, Phys. Lett. \textbf{B 271,}
61 (1991).

\bibitem {numerical1}N.K. Glendenning, T. Kodama and F.R. Klinkhamer, Phys.
Rev. D \textbf{38,} 3226 (1988); B. M. A. G. Piette and G. I. Probert, Phys.
Rev.D \textbf{75,} 125023 (2007); G.W. Gibbons, C. M. Warnick and W. W. Wong,
J. Math. Phys. \ \textbf{52,} 012905 (2011); S. Nelmes and B. M. Piette, Phys.
Rev. D \textbf{84,} 085017 (2011).

\bibitem {numerical2}P. Bizon and T. Chmaj,\ Phys. Rev.\ D \textbf{58}, 041501
(1998); P. Bizon, T. Chmaj and A. Rostworowski, Phys. Rev. D \textbf{75,}
121702 (2007); S. Zajac, Acta Phys.Polon. B \textbf{40}, 1617 (2009); S.
Zajac, Acta Phys. Polon. B \textbf{42,} 249 (2011).

\bibitem {cosmo}F. Canfora, A. Giacomini and S. A. Pavluchenko, Phys. Rev.
\textbf{D 90}, 043516 (2014).

\bibitem {cosmo2}L. Parisi, N. Radicella and G. Vilasi, Phys.Rev. \textbf{D
91}, 063533 (2015).

\bibitem {cosmo3}A. Paliathanasis and M. Tsamparlis, J. Geom. Phys.
\textbf{114}, 1 (2017).

\bibitem {canfora}F. Canfora and P.~Salgado-Rebolledo, Phys. Rev.
\textbf{D87}, 045023 (2013).

\bibitem {canfora2}F. Canfora and H.~Maeda, Phys. Rev. D \textbf{87}, 084049 (2013).

\bibitem {canfora3}F. Canfora and Phys. Rev. D \textbf{88}, 065028 (2013).

\bibitem {canfora4}F. Canfora, F. Correa and J. Zanelli, Phys. Rev. \textbf{D
90}, 085002 (2014).

\bibitem {yang1}S. Chen, Y. Li and Y. Yang, Phys. Rev. \textbf{D 89,} 025007 (2014).

\bibitem {canfora5}F. Canfora, M. Kurkov, M. Di Mauro and A. Naddeo,
Eur.Phys.J. \textbf{C 75,} 443 (2015).

\bibitem {yang2}S. Chen and Y. Yang, Nucl. Phys. \textbf{B 904}, 470 (2016).

\bibitem {canfora6}E. Ayon-Beato, F. Canfora and J. Zanelli, Phys. Lett.
\textbf{B 752}, 201 (2016).

\bibitem {canfora7}F. Canfora and G. Tallarita, Phys. Rev. \textbf{D 94,}
025037 (2016).

\bibitem {sergei}S.A. Pavluchenko, Phys. Rev. \textbf{D 94}, 044046 (2016).

\bibitem {curved}N. S. Manton and P. J. Ruback, Phys. Lett. \textbf{B 181},
137 (1986). N. S. Manton, Comm. Math. Phys. \textbf{111}, 469 (1987).

\bibitem {bratek}L.Bratek, Nonlinearity \textbf{16}, 1539 (2003).

\bibitem {Ermakov}V. Ermakov Univ. Isz. Kiev Series III\textbf{ 9}, (1880)
(translated by Harin A O).

\bibitem {Pinney}E. Pinney, Proc. Am. Math. Soc. \textbf{1,} 681 (1950).

\bibitem {ErmakovA}P.G.L. Leach and K. Andriopoulos, Appl. Anal. Discrete
Math. \textbf{2,} 146 (2008).

\bibitem {ErmakovB}F.L. Williams and P.G. Kevrekidis, Class.\ Quantum. Gravit.
\textbf{20,} L177 (2003).

\bibitem {ErmakovC}M. Tsamparlis and A. Paliathanasis, J. Phys. A: Math.
Theor. \textbf{45,} 275202 (2012).

\bibitem {Eins1}J.D. Barrow, G. Ellis, R. Maartens and C. Tsagas, Class.
Quant. Grav. \textbf{20,} L155-L164 (2003).

\bibitem {Eins2}J.D.\ Barrow and K. Yamamoto, Phys. Rev. \textbf{D 85,} 083505 (2012).

\bibitem {arnold}V.I. Arnol'd, Mathematical Methods of Classical Mechanics,
Graduate Texts in Mathematics Vol. 60, Springer, New-York (1989).

\bibitem {Haw}C.B. Collins and S.W. Hawking, Ap. J. \textbf{180,} 317 (1973).
\end{thebibliography}
\end{document}